\documentclass[aps, prb, notitlepage, citeautoscript, superscriptaddress, eqsecnum, reprint]{revtex4-2}

\usepackage{xr}
\usepackage{amsmath}
\usepackage{amsfonts, amssymb,amsxtra}
\usepackage[]{graphicx}
\usepackage{grffile}
\usepackage{amsfonts}
\usepackage{framed}
\usepackage{bbm}
\usepackage{braket}
\usepackage{xcolor}
\usepackage{mathtools}
\usepackage{LatexCommands}

\usepackage[colorlinks=true]{hyperref}
\hypersetup{
    unicode=false,          
    pdftoolbar=true,        
    pdfmenubar=true,        
    pdffitwindow=false,     
    pdfstartview={FitH},    
    pdftitle={},    
    pdfauthor={},     
    pdfsubject={},   
    pdfcreator={},   
    pdfproducer={}, 
    pdfkeywords={} {} {}, 
    pdfnewwindow=true,      
    colorlinks=true,       
    linkcolor=magenta, 
    citecolor=blue,        
    filecolor=magenta,      
    urlcolor=blue           
}

\begin{document}
\title{Error-Mitigated Simulation of Quantum Many-Body Scars on Quantum Computers with Pulse-Level Control}

\author{I-Chi Chen}
\affiliation{Department of Physics and Astronomy, Iowa State University, Ames, Iowa 50011, USA}

\author{Benjamin Burdick}
\affiliation{Department of Physics and Astronomy, Iowa State University, Ames, Iowa 50011, USA}

\author{Yongxin Yao}
\affiliation{Department of Physics and Astronomy, Iowa State University, Ames, Iowa 50011, USA}
\affiliation{Ames National Laboratory, Ames, Iowa 50011, USA}

\author{Peter P.~Orth}
\email{porth@iastate.edu}
\affiliation{Department of Physics and Astronomy, Iowa State University, Ames, Iowa 50011, USA}
\affiliation{Ames National Laboratory, Ames, Iowa 50011, USA}

\author{Thomas Iadecola}
\email{iadecola@iastate.edu}
\affiliation{Department of Physics and Astronomy, Iowa State University, Ames, Iowa 50011, USA}
\affiliation{Ames National Laboratory, Ames, Iowa 50011, USA}

\begin{abstract}
Quantum many-body scars are an intriguing dynamical regime in which quantum systems exhibit coherent dynamics and long-range correlations when prepared in certain initial states. We use this combination of coherence and many-body correlations to benchmark the performance of present-day quantum computing devices by using them to simulate the dynamics of an antiferromagnetic initial state in mixed-field Ising chains of up to 19 sites. In addition to calculating the dynamics of local observables, we also calculate the Loschmidt echo and a nontrivial unequal-time connected correlation function that witnesses long-range many-body correlations in the scarred dynamics. We find coherent dynamics to persist over up to 39 Trotter steps even in the presence of various sources of error. To obtain these results, we leverage a variety of error mitigation techniques including noise tailoring, zero-noise extrapolation, dynamical decoupling, and physically motivated postselection of measurement results. Crucially, we also find that using pulse-level control to implement the Ising interaction yields a substantial improvement over the standard controlled-NOT-based compilation of this interaction. Our results demonstrate the power of error mitigation techniques and pulse-level control to probe many-body coherence and correlation effects on present-day quantum hardware.
\end{abstract}
\date{\today}

\maketitle

Quantum simulation is one of the most natural problems in which to expect a quantum computer to outperform a classical one. In particular, simulating the time evolution of a quantum many-body system with local interactions on a quantum computer requires a number of local quantum gates scaling polynomially with the number $N$ of qubits and linearly with the total simulation time $T$~\cite{Lloyd96}. Classical simulation techniques, in contrast, require resources scaling exponentially with $N$, restricting most studies of quantum many-body dynamics to small systems and/or early times. There is thus a relatively clear path to quantum advantage for the quantum simulation problem, assuming the availability of quantum hardware that can simulate the dynamics of a quantum many-body system with sufficiently large $N$ and $T$.

The present generation of quantum hardware operates in the so-called noisy intermediate-scale quantum (NISQ) regime~\cite{Preskill18}. NISQ devices have enough qubits ($\sim 10^1$-$10^3$) to potentially evade classical simulability, but coherence times and gate fidelities that are too small to simulate dynamics beyond the early-time regime, where tensor-network methods~\cite{Schollwock11} are often still applicable. As hardware continues to improve, a major challenge of the NISQ era is to determine strategies to maximize the utility of such devices despite their imperfections. One approach to this problem is to use variational quantum algorithms~\cite{Cerezo21,Bharti22}, which reduce the required circuit depth for tasks including time evolution~\cite{Li17,Yuan19, Yao-AVQDS-PRX_Q-2021,barrattParallelQuantumSimulation2021, Barison21,Lin21, benedettiHardwareefficientVariationalQuantum2021, mansurogluClassicalVariationalOptimization2021,Berthusen21} at the cost of requiring many circuit evaluations. An alternative approach is to apply a growing toolbox of hardware- and noise-aware quantum error mitigation techniques to a relatively simple quantum simulation algorithm, e.g.~the first-order Trotter method of Ref.~\cite{Lloyd96}. Error mitigation strategies can be applied at the hardware level, e.g. by applying dynamical decoupling pulses to idle qubits~\cite{Lorenza98,Pokharel18,Jurcevic2021} or designing optimized pulse sequences to reduce gate execution times~\cite{Stenger2021,Kim21}. Other strategies, such as randomized compilation~\cite{Wallman16,Li17} and zero noise extrapolation~\cite{Li17,Temme17,Kandala19,Giurgica-Tiron-ZNE-2020}, run additional circuits that are logically equivalent to the target circuit and perform postprocessing of the data to estimate the hypothetical noiseless result. Recently, Ref.~\cite{Kim21} showed that combining these error-mitigation techniques yields first-order Trotter simulations of quantum dynamics whose accuracy as measured by local observables is competitive with tensor-network methods, at least for low bond dimensions.

In this work, we test the ability of present-day quantum hardware to simulate nontrivial many-body dynamics and correlation effects in a prototypical interacting quantum system: the one-dimensional mixed-field Ising model (MFIM). This model and its variants have been simulated on NISQ hardware in several recent works~\cite{Sopena21,vovroshConfinementEntanglementDynamics2021,Frey_Rachel-ScienceAdvances-2022}. Our goal here is to model a particular physical phenomenon known as quantum many-body scars (QMBS)~\cite{Turner18a,Moudgalya18a,Turner18b,Moudgalya18b} (see Refs.~\cite{Serbyn21,Moudgalya22,Chandran22} for reviews). This phenomenon was observed experimentally in an analog quantum simulation of the MFIM using Rydberg atoms in optical tweezers~\cite{Bernien17}, where coherent oscillations of the local Pauli expectation values $\braket{Z_i(t)}$ were observed after a quantum quench from the N\'eel state $\ket{Z_2}=\ket{010\dots}$ or its $\mathbb Z_2$ conjugate $\ket{Z'_2}=\ket{101\dots}$. This came as a surprise, since the MFIM is known to be nonintegrable for generic values of the transverse and longitudinal fields and the N\'eel states $\ket{Z_2}$ and $\ket{Z'_2}$ have a finite energy density relative to the ground state of the model. In such a case, reasoning based on the eigenstate thermalization hypothesis (ETH)~\cite{Deutsch91,Srednicki94} leads to the expectation that the dynamics from these initial states should rapidly decohere~\cite{Rigol08,D'Alessio16}. Intriguingly, the oscillatory dynamics also encode coherent oscillations in space: Ref.~\cite{Iadecola19} argued based on finite-size numerics that the dynamics from the N\'eel state exhibits long-range \textit{connected} unequal-time correlations at wavenumber $\pi$. This finding also defies intuition based on the ETH and suggests that the oscillatory dynamics observed in the experiment \cite{Bernien17} are fundamentally many-body in nature, and cannot be explained by the precession of free spins. QMBS are known to occur in a variety of other models~\cite{Moudgalya18a,Schecter19,Bull19a,Hudomal19,Iadecola20,Mark20,Moudgalya20,Pakrouski20,ODea20,Ren21,Tang21}, and have been observed in several analog quantum simulation experiments~\cite{Bernien17,Su22,Zhang22}, but we focus here on a digital quantum simulation approach.

Motivated by these experimental and theoretical results, we use IBM quantum processing units (QPUs) to perform a first-order Trotter simulation of the MFIM in the regime with QMBS for up to 39 Trotter steps on systems of up to 19 qubits. One goal of the study is to use the coherent dynamics in the scarred regime to benchmark the performance of these QPUs; another is to use QPUs to verify the presence of nontrivial connected unequal-time correlations in the dynamics of the N\'eel state. Such correlations, which demonstrate the inextricably many-body nature of the oscillatory dynamics, are challenging to measure in analog quantum simulators and have yet to be probed experimentally. Our QPU results demonstrate that they can be accessed using digital quantum simulation.

To optimize the performance of the devices, we make use of a variety of techniques. First, we implement quantum simulation of the Ising interaction using a scaled cross-resonance pulse and show that this implementation outperforms the naive compilation of the Ising evolution operator using two controlled-NOT (CNOT) gates. Second, we apply an arsenal of error mitigation techniques, including zero-noise extrapolation, Pauli twirling, dynamical decoupling, readout error mitigation, and, where appropriate, physically motivated postselection of computational-basis measurement outcomes. Many of these methods were applied to simulate the transverse-field Ising model on the heavy hexagon lattice for at most 20 Trotter steps in Ref.~\cite{Kim21}; we will comment on areas where our implementation differs from theirs, the most important being our use of postselection and a new approach to Pauli twirling of non-Clifford gates.

We combine these techniques to calculate spatially averaged local observables, as well as more sensitive probes of the dynamics including the Loschmidt echo and a nontrivial finite-wavenumber connected correlation function. We find that combining pulse level control and error mitigation techniques extends by roughly a factor of two the timescales over which nontrivial oscillatory dynamics can be observed. Taken together, our results demonstrate that these nontrivial many-body effects can be probed, at least in the early time regime, on present-day quantum hardware.

The remainder of the paper is organized as follows. In Sec.~\ref{sec:Model and Observables} we define the mixed-field Ising model and the related observables that we will calculate on the QPU. In Sec.~\ref{sec:Pulse-Level Implementation of the Ising Interaction}, we describe the scaled cross-resonance pulse (``scaled-$R_{ZX}$" for short) implementation of the Ising interaction. We present data benchmarking its performance against the more standard implementation of the interaction using two CNOT gates. In Sec.~\ref{sec:Error Mitigation Techniques}, we present data for simulations of a 12- and 19-site chain using both the two-CNOT and scaled-$R_{ZX}$ implementations, and use this to motivate a brief discussion of the error mitigation techniques we use. We show our main results in Sec.~\ref{sec:Error-Mitigated Results}, which includes error-mitigated results for the staggered magnetization, the Loschmidt echo, and the finite-wavenumber connected unequal-time correlator. Finally, conclusions and outlook are discussed in Sec.~\ref{sec:Conclusions and Outlook}.

\section{Model and Observables} 
\label{sec:Model and Observables}
\subsection{Model}

In this work we simulate the dynamics of a chain of $L$ spin-1/2 degrees of freedom generated by the Hamiltonian 
\begin{align}
\label{eq:HRydberg}
H = 4V\sum^{L-1}_{i=1} n_in_{i+1}+\Omega\sum^L_{i=1} X_i,
\end{align}
where $n_i=\frac{I-Z_i}{2}$, and where $Z_i$ and $X_i$ are Pauli operators on site $i$. This model arises when studying chains of trapped Rydberg atoms with rapidly decaying van der Waals interactions, where the operator $n_i$ is interpreted as an occupation number for the local atomic Rydberg state and the operator $X_i$ induces transitions between the ground and Rydberg state. In this paper, we will label computational basis (CB) states using the eigenstates $\ket{0}_i$ and $\ket{1}_i$ of the $Z_i$ operator for which $n_i\ket{0}_i = 0$ and $n_i \ket{1}_i = \ket{1}_i$.
Rewriting Eq.~\eqref{eq:HRydberg} in terms of Pauli matrices, we obtain
\begin{align}
\label{eq:H}
    &H=H_{ZZ}+H_{Z}+H_{X}\\
    &=V\sum^{L-1}_{i=1} Z_iZ_{i+1}-2V\sum^{L-1}_{i=2}Z_i-V(Z_1+Z_L)+\Omega\sum^L_{i=1}X_i.\nonumber
\end{align}
The Hamiltonian \eqref{eq:H} is an Ising model with transverse and longitudinal fields (the MFIM), but in which the strength of the longitudinal field is tied to the interaction strength $V$ as a consequence of the model's origin in Eq.~\eqref{eq:HRydberg}. Note that the model is written with open boundary conditions, and that the longitudinal field strength is reduced by a factor of two on the first and last sites of the chain $(i=1, L)$.

The model \eqref{eq:H} is nonintegrable for generic values of $V$ and $\Omega$. QMBS emerge in the limit $V\gg \Omega$, which is known in the Rydberg-atom literature as the ``Rydberg blockade" regime~\cite{Jaksch00,Lukin01}. In this limit, computational basis states with different eigenvalues of $\sum^{L-1}_{i=1} n_{i}n_{i+1}$ decouple into sectors separated by an energy scale $\sim V$. When the system is initialized in a CB state with $\braket{\sum^{L-1}_{i=1} n_{i}n_{i+1}}=0$, such as the N\'eel states $\ket{Z_2}$ and $\ket{Z'_2}$, the probability of finding nearest-neighbor sites in the configuration $\ket{11}$ is heavily suppressed. The subspace of the full Hilbert space in which no two nearest-neighbor sites are in the state $\ket{11}$ is known as the Fibonacci Hilbert space, as the number of such states scales as $\varphi^L$, where $\varphi$ is the golden ratio. An effective model for the system in this limit is known as the ``PXP model," which arises from the projection of $H_X$ into the Fibonacci Hilbert space~\cite{Bernien17,Turner18a}.

\begin{figure}[t]
    \centering
    \includegraphics[width=\columnwidth]{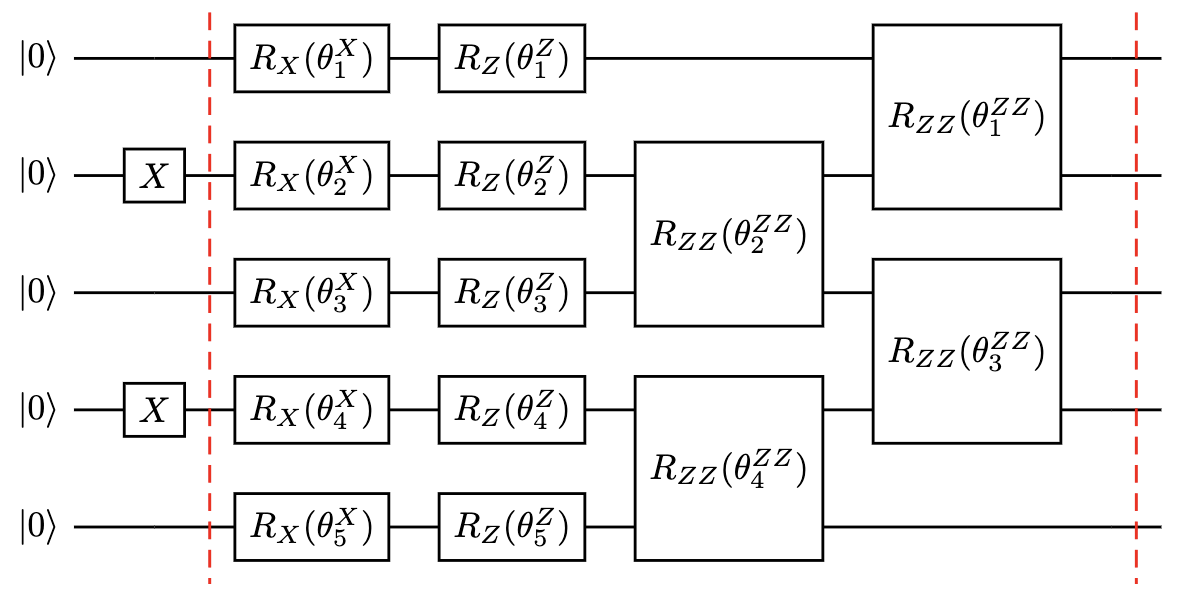}
    \caption{Schematic of the state-preparation circuit and the Trotter circuit for one time step $\Delta t$ in an $L=5$ site chain. The first part of the circuit prepares the N\'eel state $\ket{01010}$ from the polarized state $\ket{00000}$ using $X$ gates. The Trotter circuit uses single-qubit rotations $R_{X}(\theta^X_i)=e^{-i\theta^X_i X_i/2}$ and $R_{Z}(\theta^Z_i)=e^{-i\theta^Z_i Z_i/2}$ and two-qubit gates $R_{ZZ}(\theta^{ZZ}_{i})=e^{-i\theta^{ZZ}_{i}Z_{i}Z_{i+1}/2}$. Here, $\theta^{Z}_1 = \theta^{Z}_{L} = -2V\Delta t$. The remaining angles are given by $\theta^X_i = 2\Omega \Delta t$, $\theta^Z_i = -4V \Delta t$, and $\theta^{ZZ}_i = 2V\Delta t$ for $i = 1,2,\ldots, L$.}
    \label{fig:trotter_1step}
\end{figure}

To simulate the system's dynamics under the Hamiltonian \eqref{eq:H}, we employ a first-order Trotter decomposition of the unitary evolution operator over a time $\Delta t$:
\begin{equation}
\label{eq:Trotter}
    U(\Delta t) \approx e^{-iH_{ZZ}\Delta t} e^{-iH_{Z}\Delta t} e^{-iH_{X}\Delta t}.
\end{equation}
A decomposition of this circuit into single-qubit gates $R_{X}(\theta^X_i)=e^{-i\theta^X_i X_i/2}$ and $R_{Z}(\theta^Z_i)=e^{-i\theta^Z_i Z_i/2}$ and two-qubit gates $R_{ZZ}(\theta^{ZZ}_{i})=e^{-i\theta^{ZZ}_{i}Z_{i}Z_{i+1}/2}$ is shown in Fig.~\ref{fig:trotter_1step}. Evolution over a time $T=n\Delta t$ is obtained by applying the circuit \eqref{eq:Trotter} $n$ times.

\subsection{Observables}
To probe the dynamics of the model \eqref{eq:H} in the QMBS regime $V\gg\Omega$, we use the QPU to measure three dynamical properties. First, to characterize the oscillations, we measure the dynamics of the expectation value of the staggered magnetization operator,
\begin{align}
\label{eq:Zpi}
    Z_\pi = \sum^L_{i=1}(-1)^{i}Z_i.
\end{align}
This operator takes its extremal values $\braket{Z_\pi}=\pm L$ when the system is prepared in the N\'eel state $\ket{Z'_2}$ or $\ket{Z_2}$, respectively. Exact simulations of a quantum quench from the $\ket{Z_2}$ state show a weakly damped coherent oscillation of $\braket{Z_\pi(t)}$ with frequency $\omega\approx 1.33\, \Omega$~\cite{Turner18a}, indicating that the system's state is periodically cycling between the two N\'eel states. In contrast, the expectation based on the ETH is that $\braket{Z_\pi(t)}$ would decay rapidly, on a timescale $\sim 1/\Omega$, to its thermal value of $0$.
These coherent oscillations arise due to the presence of a tower of eigenstates with roughly equal energy spacings in the many-body spectrum~\cite{Turner18a}. These ``scar states" have high overlap with the N\'eel states, so preparing the system in one of these initial states projects the ensuing dynamics strongly onto this set of special eigenstates. $\braket{Z_\pi(t)}$ is calculated on the QPU by performing Trotter evolution out to time $t$ and measuring the state in the computational basis. 

\begin{figure*}[t]
\includegraphics[width=2.0\columnwidth]{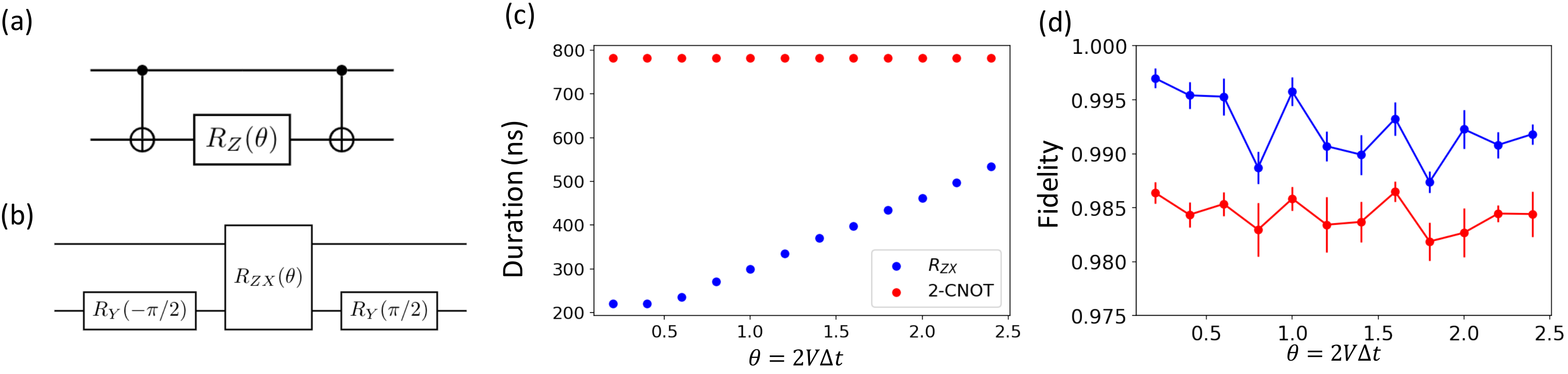}
\centering
\caption{
Benchmarking the two-CNOT and scaled $R_{ZX}$ implementations of the $R_{ZZ}(\theta)$ gate. All data were taken on the IBM QPU Casablanca (\texttt{ibmq\_casablanca}).
(a) Standard implementation of the $R_{ZZ}(\theta)$ gate using two CNOTs.
(b) Implementation of the $R_{ZZ}(\theta)$ gate using an $R_{ZX}(\theta)$ gate and appropriate single-qubit rotations.
(c) Duration in ns of the pulse schedule on \texttt{ibmq\_casablanca} as a function of $\theta = 2V\Delta t\leq 2.5$ for the two-CNOT and scaled-$R_{ZX}$ implementations of $R_{ZZ}(\theta)$. The scaled-$R_{ZX}$ implementation has a substantially shorter pulse duration for all $\theta$ considered.
(d) Fidelity of the two-CNOT and scaled-$R_{ZX}$ implementations of $R_{ZZ}(\theta)$ obtained using quantum process tomography on \texttt{ibmq\_casablanca} as described in the text. The fidelity of the scaled-$R_{ZX}$ implementation decreases with increasing $\theta$ while that of the two-CNOT implementation remains roughly constant. Data were accumulated over two days, so fluctuations in the fidelity due to calibration drifts are visible.
}
\label{fig:Rzz}
\end{figure*}

Second, we measure the Loschmidt echo,
\begin{align}
\label{eq:L}
    \mathcal L(t) = |\braket{\psi(t)|\psi(0)}|^2,
\end{align}
where $\ket{\psi(0)}$ is taken to be $\ket{Z_2}$. When the system is prepared in this initial state, the scarred eigenstates give rise to sharp periodic revivals of the Loschmidt echo to a value of order 1, with period matching that of the oscillations in $\braket{Z_\pi(t)}$~\cite{Turner18b}. This behavior is highly atypical---$\mathcal L(t)$ is expected to decay to zero exponentially fast in quantum quenches of nonintegrable models from typical high-energy-density initial states~\cite{Gorin06,Goussev16}. Thus, the Loschmidt echo is a much more sensitive quantity than $\braket{Z_\pi(t)}$, which is built from expectation values of local observables. To measure the Loschmidt echo on the QPU, we perform Trotter evolution of the $\ket{Z_2}$ state and measure in the CB to determine the probability to be in the state $\ket{Z_2}$ after a time $t$.

Third, we measure the correlation function
\begin{subequations}
\label{eq:CY}
\begin{align}
    \mathcal C_Y(t)=\braket{Y_{\pi}(t)Y_{\pi}(0)},
\end{align}
where
\begin{align}
\label{eq:Ypi}
    Y_{\pi}\equiv\sum^{L}_{i=1}\left(-1\right)^{i}(PYP)_i,
\end{align}
with
\begin{align}
\label{eq:PYP_def}
    (PYP)_{i}=\begin{cases}
    Y_1 P_2 & i=1\\
    P_{i-1}Y_iP_{i+1}& i=2,\dots,L-1\\
    P_{L-1}Y_L & i=L
    \end{cases}
\end{align}
\end{subequations}
and $P_{i}=(1+Z_i)/2$ such that $P_i|0\rangle_i = |0\rangle_i$ and $P_i|1\rangle_i = 0$. 
Note that when $\mathcal C_Y(t)$ is evaluated in the state $\ket{Z_2}$ (or indeed any other initial CB state), the disconnected part of the correlator vanishes since $\braket{Z_2|Y_\pi(0)|Z_2}=0$. Thus, $\mathcal C_Y(t)$ probes nontrivial long-range correlations in space and time with wavenumber $\pi$. In Ref.~\cite{Iadecola19}, it was argued that $\mathcal C_Y(t)$ exhibits coherent weakly damped oscillations when the initial state is taken to be one of the N\'eel states. These long-range correlations arise due to the presence of off-diagonal long-range order in the scarred eigenstates \cite{Yang62,Iadecola19}. This nontrivial correlator has yet to be measured experimentally. Measuring it on a QPU is challenging, but achievable using the ancilla-free protocol of Ref.~\cite{Mitarai19}, which we describe further in Sec.~\ref{sec:Corr}.

\section{Pulse-Level Implementation of the Ising Interaction}
\label{sec:Pulse-Level Implementation of the Ising Interaction}

Implementing the Trotter circuit in Eq.~\eqref{eq:Trotter} requires realizing the two qubit gate $R_{ZZ}(\theta)$ on the device. One approach to solving this problem is to decompose $R_{ZZ}(\theta)$ into a basis gate set, the most standard of which includes CNOT and arbitrary one-qubit rotations. In this basis, $R_{ZZ}(\theta)$ can be realized by applying two CNOT gates on either side of an $R_Z(\theta)$ gate on the target qubit~\cite{Vatan04,Smith2019}, as shown in Fig.~\ref{fig:Rzz}(a).

An alternative approach is to leverage knowledge of the basic set of pulses used to generate two-qubit gates at the hardware level. On IBM QPUs, well-calibrated CNOT gates are built by adding single qubit gates before and after the $R_{ZX}(\pi/2)$ gate \cite{Alexander20}, which is defined via $R_{ZX}(\theta)=e^{-i\theta Z_{c}X_{t}/2}$, where $c$ and $t$ denote the control and target qubits, respectively. These $R_{ZX}(\pi/2)$ gates are realized using an echoed cross resonance pulse sequence described in further detail in Appendix~\ref{app:Mitigation} and in Ref.~\cite{Alexander20}. The ability to implement an $R_{ZX}(\theta)$ gate opens another route to realize the $R_{ZZ}(\theta)$ gate simply by dressing the $R_{ZX}(\theta)$ gate with $R_{Y}(\pm\pi/2)$ gates on the target qubit, see Fig.~\ref{fig:Rzz}(b). To realize the $R_{ZX}(\theta)$ gate with arbitrary rotation angle on the QPU, we scale the pulse amplitudes and durations used to generate the $R_{ZX}(\pm\pi/2)$ gate in the manner described in Refs.~\cite{Stenger2021,Earnest21} and summarized in Appendix~\ref{app:Mitigation}. We therefore refer to this pulse-level implementation of the $R_{ZZ}$ gate as the ``scaled-$R_{ZX}$" implementation. The pulse sequences are programmed using Qiskit pulse~\cite{Qiskit,Alexander20}; examples of pulse schedules used in our simulations are shown in Appendix~\ref{app:Mitigation} (Fig.~\ref{fig:Schedule}).

Since the scaled-$R_{ZX}$ approach uses fewer cross-resonance pulses than the two-CNOT implementation of $R_{ZZ}$, we expect the former to yield pulse schedules with shorter overall duration than the latter. The duration of the pulse schedule that realizes the two-CNOT implementation of $R_{ZZ}(\theta)$ is independent of the rotation angle $\theta$, since the $R_{Z}(\theta)$ gate is simply realized as a phase shift on the pulse schedule~\cite{Alexander20}. 
In contrast, in the scaled-$R_{ZX}$ implementation of $R_{ZZ}$, the cross-resonance pulse duration depends roughly linearly on $\theta$ for $\theta$ above a certain threshold and is constant below that threshold (see Appendix~\ref{app:Mitigation}). The pulse durations in ns of the 2-CNOT and scaled-$R_{ZX}$ implementations of $R_{ZZ}(\theta)$ on the IBM QPU Casablanca (\texttt{ibmq\_casablanca}) are shown as a function of $\theta$ in Fig.~\ref{fig:Rzz}(c). Despite the $\theta$-dependence of the scaled-$R_{ZX}$ pulse duration, there is a wide range of interaction strengths $V$ and Trotter time steps $\Delta t$ giving an angle $\theta=2V\Delta t$ such that the scaled-$R_{ZX}$ implementation has the shorter pulse duration of the two methods.

The total duration of the pulse schedule that realizes a given quantum gate is positively correlated with the gate's error rate. We therefore expect that the scaled-$R_{ZX}$ implementation of $R_{ZZ}$ should have a lower error rate than that of the two-CNOT implementation of the same gate.
To compare the error rates of the $R_{ZZ}$ gates rates realized using the two approaches, we measure the fidelity of $R_{ZZ}(\theta)$ at 12 different angles from $\theta=0.2$ to $\theta=2.4$ using quantum process tomography (QPT)~\cite{O'Brien04,Garion21}, which is built into IBM's Ignis module, with state preparation basis $\{\ket{0}, \ket{1}, \ket{X_+}, \ket{Y_+}\}$ and measurement basis $\{X,Y,Z\}$ for each qubit. In order to obtain an estimate of the gate error that is decoupled from state preparation and measurement (SPAM) errors, we use a simple scheme relying on gate folding. Letting $G=R_{ZZ}(\theta)$, we consider the sequence of logically equivalent gates $G$, $G G^\dagger G$, and $G G^\dagger G G^\dagger G$, which correspond to a ``scale factor" of $\lambda=1, 3,$ and $5$, respectively. For each $\lambda$, we use QPT to estimate the average gate fidelity, which includes SPAM errors. This fidelity decreases with $\lambda$ because gate folding increases the gate noise in the circuit. We then fit the resulting data points to a linear model $F_0-\epsilon \lambda$, which is justified under the assumptions that $\epsilon$ is small and that $G$ and $G^\dagger$ have identical error rates.
The slope $\epsilon$ is an estimate of the error rate that is free of SPAM errors, since these errors do not scale with $\lambda$. To obtain the results plotted in Fig.~\ref{fig:Rzz}(d), we ran the above procedure on the IBM QPU Casablanca using qubits $q1$ and $q3$, with 1024 shots for each measurement and using complete readout error mitigation~\cite{Bravyi21} (see also Appendix~\ref{app:Mitigation}). We also repeated QPT four times for each scale factor to collect better statistics; the linear fit to extract $\epsilon$ was performed over the full data set. The error bars for each $\theta$ value represent the standard deviation of the slope calculated from the covariance matrix of the linear fit for that $\theta$. 

Fig.~\ref{fig:Rzz}(d) shows that the fidelity of the scaled-$R_{ZX}$ implementation of $R_{ZZ}(\theta)$ decreases with increasing $\theta$, while the fidelity of the two-CNOT implementation remains almost constant. Because the experiment was implemented across two days, there are also fluctuations in the fidelity due to calibration drifts (see also Ref.~\cite{Stenger2021}). At the smallest value of $\theta=0.2$, our fidelity results indicate that the scaled-$R_{ZX}$ approach realizes an $\sim80\%$ reduction in the error rate of the $R_{ZZ}(\theta)$ gate relative to the two-CNOT implementation. This decreases to a $\sim50\%$ error reduction at the largest value of $\theta=2.5$. This is consistent with the pulse duration results plotted in Fig.~\ref{fig:Rzz}(c), which show that the pulse durations of the two-CNOT and scaled-$R_{ZX}$ implementations approach one another with increasing $\theta$.

We now comment on our choice of Hamiltonian parameters $V$ and $\Omega$ for simulating the system's dynamics in the regime with QMBS. As mentioned in Sec.~\ref{sec:Model and Observables}, we need $V\gg \Omega$ in order to be in the QMBS regime. However, according to Fig.~\ref{fig:Rzz}(c), reducing $V$ yields a shorter pulse duration for the scaled-$R_{ZX}$ implementation of $R_{ZZ}$, thereby reducing the gate error. Moreover, $\Omega$ determines the frequency of the oscillations that are characteristic of QMBS, so choosing a larger $\Omega$ is desirable in order to manifest more oscillation periods within a fixed time window. The choice of $\Delta t$ is essential as well, since the Trotter error scales to leading order as $V\Omega (\Delta t)^2$. After trying many sets of parameters, we settled on $V=1$, $\Omega=0.24$, and $\Delta t = 1$ as optimal parameters to simulate the system's dynamics in the QMBS regime. These parameters correspond to an $R_{ZZ}$ rotation angle $\theta=2.0$, where the data in Fig.~\ref{fig:Rzz}(d) indicate that the scaled-$R_{ZX}$ approach yields a $\sim 57\%$ error reduction relative to the two-CNOT implementation. Although the sizable value of $\Delta t$ incurs substantial Trotter error (see Appendix~\ref{app:Mitigation} for a comparison between Trotter and exact dynamics), the Trotter circuit with $\Delta t=1$ nevertheless exhibits pronounced coherent oscillations with period $2\pi/(1.33\Omega)\approx 19.68$ for this choice of parameters. For this choice of parameters, the dimensionless quantity $Vt$ is simply the number of Trotter steps. 

We note in passing that our Trotter circuit \eqref{eq:Trotter} could alternatively be viewed as a Floquet circuit due to the relatively large value of $\Delta t$. That is, we can view this circuit as simulating not the dynamics under the time-independent Hamiltonian in Eq.~\eqref{eq:H}, but rather the time-dependent Hamiltonian
\begin{align}
\label{eq:Floquet}
    H(t) = f_+(t)\, H_X + f_-(t)\, (H_{ZZ}+H_{Z}),
\end{align}
where $f_\pm(t)=[1\pm\text{sgn}(\sin \frac{2\pi}{\Delta t} t)]/2$, whose evolution operator over a time $\Delta t$ is precisely the right-hand side of Eq.~\eqref{eq:Trotter}.
QMBS have been studied in several Floquet variants of the mixed-field Ising and PXP models, see e.g. Refs.~\cite{Iadecola20b,Mukherjee20,Bluvstein21,Maskara21,Hudomal22}, so it is by now well established that they can exist in this periodically driven setting. Our results demonstrate their existence in the model \eqref{eq:Floquet}.

\section{Error Mitigation Techniques}
\label{sec:Error Mitigation Techniques}

\begin{figure}[t]
\includegraphics[width=1.0\columnwidth]{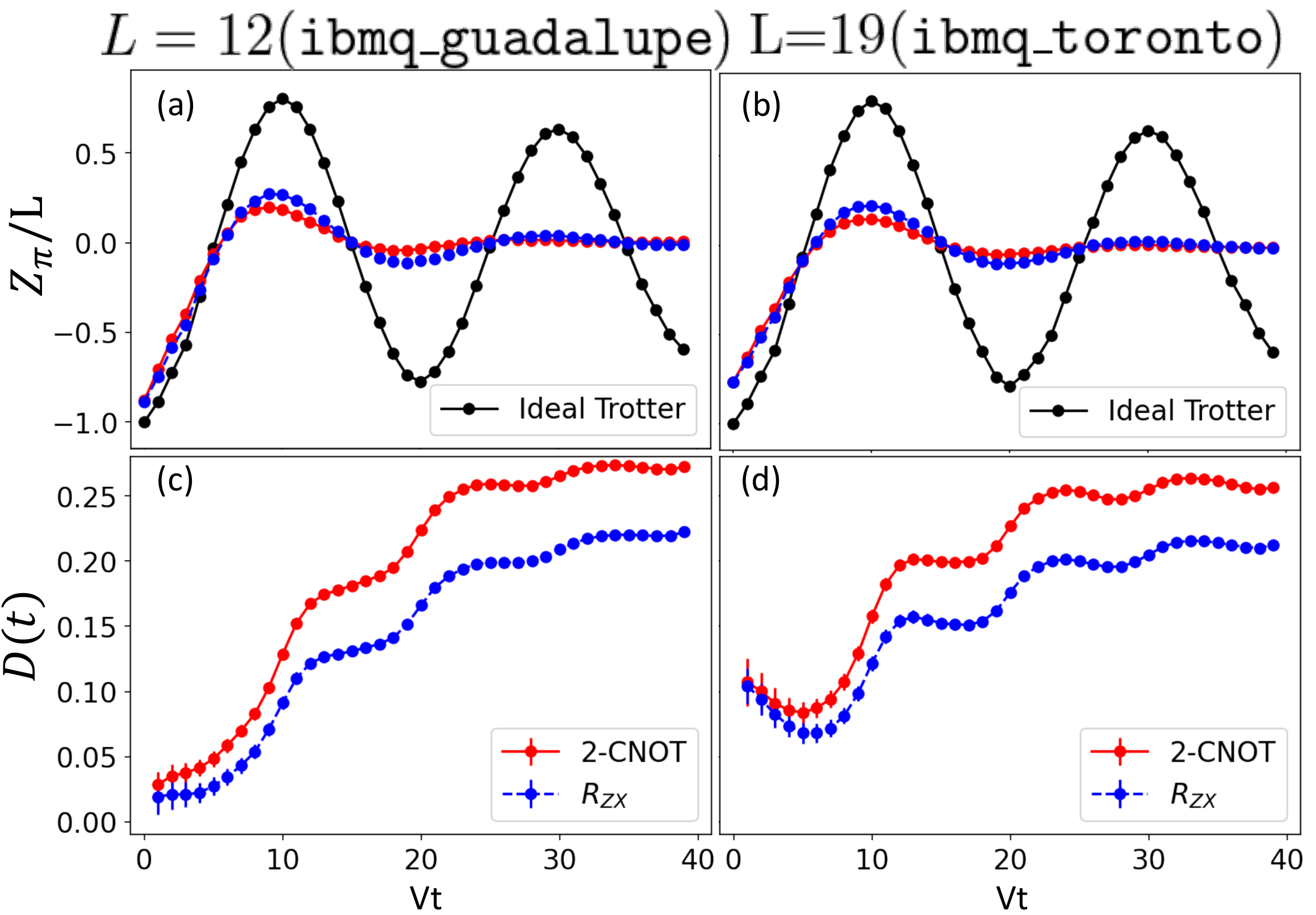}
\centering
\caption{
Unmitigated Trotter simulation of the staggered magnetization density $\braket{Z_\pi(t)}/L$ [(a),(b); see Eq.~\eqref{eq:Zpi}] and accumulated error $D(t)$ [(c),(d); see Eq.~\eqref{eq:Dt}] from the initial state $\ket{Z_2}$. Data for chains of 12 [(a),(c)] and 19 [(b),(d)] qubits were obtained using \texttt{ibmq\_guadalupe} and \texttt{ibmq\_toronto}, respectively. Data for both the two-CNOT (red) and scaled-$R_{ZX}$ (blue) implementations of the Trotter circuit are shown, with noiseless Trotter simulation results (black) for reference. No error mitigation was used here to directly compare the two $R_{ZZ}$ implementations. 
Error bars for each point in (a) and (b) represent the standard deviation of the data over 20 trials. Error bars in (c) and (d) are calculated from those in (a) and (b) by propagation of errors. Oscillations of $\braket{Z_\pi(t)}/L$ over roughly one period are observed for the scaled-$R_{ZX}$ implementation and are barely discernible for the two-CNOT implementation. The two-CNOT implementation also accumulates more error than the scaled-$R_{ZX}$ implementation.
}
\label{fig:Z_pi_nomit}
\end{figure}

Based on the discussion in the previous section, we expect that the scaled-$R_{ZX}$ implementation of the $R_{ZZ}$ gate will outperform the two-CNOT implementation when performing Trotter evolution of the N\'eel state $\ket{Z_2}$ under the Hamiltonian \eqref{eq:H}. We test this hypothesis by running Trotter simulations using both approaches on the IBM QPUs Guadalupe (\texttt{ibmq\_guadalupe}) and Toronto (\texttt{ibmq\_toronto}) for chains with $L=12$ and $19$ sites, respectively. For each $R_{ZZ}$ implementation, we execute 39 Trotter steps using the parameters $\Omega=0.24$, $V=1$, and $\Delta t=1$. At each time $t$, we measure $\braket{Z_i(t)}$ for all sites $i$ using 8192 shots. We repeat the time evolution procedure 20 times and average the results over these trials~\footnote{For each of the 20 evolution circuits, we use Pauli-twirled two-qubit gates as described later in this section. However, since we do not postprocess the data (e.g. by doing zero noise extrapolation), the twirling has almost no effect on the results other than to introduce some negligible single-qubit gate errors.}. To quantify the accumulation of error during the simulation, we define
\begin{align}
\label{eq:Dt}
    D(t)=\frac{1}{t}\int_{0}^{t}dt' \frac{1}{L}\sum_{i=1}^{L}\left|\left\langle Z_{i}(t')\right\rangle_{\text{SV}} -\left\langle Z_{i}(t')\right\rangle_{\text{QPU}} \right|^{2},
\end{align}
where $\left\langle Z_{i}(t)\right\rangle_{\text{SV}} $ is the result obtained from Trotter evolution on a noiseless statevector simulator and $\left\langle Z_{i}(t)\right\rangle_{\text{QPU}}$ is the result from the QPU. The results of these simulations are shown in Fig.~\ref{fig:Z_pi_nomit}. 

Figs.~\ref{fig:Z_pi_nomit}(a) and (b) show the dynamics of the staggered magnetization density $\braket{Z_{\pi}(t)}/L$ [see Eq.~\eqref{eq:Zpi}] for the 12- and 19-qubit simulations, respectively. Note that we did not apply any error mitigation techniques here in order to directly compare the two implementations of the $R_{ZZ}$ gate. For both simulations, both implementations clearly fail to reproduce the expected oscillations.
The simulation using the scaled-$R_{ZX}$ approach does slightly outperform the two-CNOT approach---in particular, one weak oscillation over a scar period $Vt\approx 20$ is barely observable for the scaled-$R_{ZX}$ data.
However, both approaches perform poorly compared to the ideal Trotter results from the statevector simulator.
For the 12-qubit system, the results begin to deviate strongly from the ideal Trotter curve after $Vt=8$. For the 19-qubit system, the calculations disagree markedly even at early times.
Figs. \ref{fig:Z_pi_nomit}(c) and (d) show that the accumulated error grows dramatically before $Vt=25$, after which it appears to saturate. The saturated value of the error is nearly the same for the 12- and 19-qubit calculations.

There are many error sources that contribute to these results.
One source is quantum thermal relaxation, whose effects are quantified by the qubit relaxation time $T_1$ and the qubit dephasing time $T_2$.
On IBM QPUs, these timescales are roughly $T_{1,2}\sim 100\ \mu\text{s}$, which limits the total circuit depth that can be executed on the device. 
Another source of error comes from imperfect operation of the physical gates. 
On IBM QPUs, single-qubit gates have error rates of $\sim10^{-4}$ to $10^{-3}$, while CNOT gates (which involve the use of $R_{ZX}$ gates as described in Sec.~\ref{sec:Pulse-Level Implementation of the Ising Interaction}) have error rates of $\sim 5\times10^{-3}$ to $1.5\times10^{-2}$. 
Thus, two-qubit gates provide the dominant source of gate error.
Finally, readout errors can occur in which the device misreports the state of a qubit as $\ket{0}$ when it is actually $\ket{1}$ and vice versa. Readout error rates can range from $\sim2\times 10^{-2}$ to $6\times 10^{-2}$, which is even larger than the CNOT error rate. However, readout only occurs once per circuit, whereas each circuit can use many two-qubit gates. (Note that readout error likely accounts for the majority of the difference between the QPU and ideal Trotter results for both the 12- and 19-qubit simulations at time $t=0$.)

The average two-qubit error rates were nearly identical for \texttt{ibmq\_guadalupe} and \texttt{ibmq\_toronto} ($1.8\times10^{-2}$ and $1.6\times10^{-2}$, respectively) when the data shown in Fig.~\ref{fig:Z_pi_nomit} were obtained. However the average readout error rate for \texttt{ibmq\_guadalupe} was $3.6\times10^{-2}$, while for \texttt{ibmq\_toronto} it was $5.4\times10^{-2}$. In fact, the highest readout error rate among the qubits we used on \texttt{ibmq\_guadalupe} was $9\times10^{-2}$, while on \texttt{ibmq\_toronto} it was $2.16\times10^{-1}$. Therefore, we believe that the poorer agreement with the ideal Trotter simulation at early times that was observed for the results obtained for the 19-qubit chain on \texttt{ibmq\_toronto} is due to the increased readout error rate on that device at the time the experiments were performed.

To reduce the impact of these various error sources, we implemented an arsenal of error mitigation techniques. 
The simplest of these techniques is readout error mitigation, which is built into Qiskit Ignis. 
We also implement dynamical decoupling using an $X_\pi-X_{-\pi}$ pulse sequence to reduce decoherence errors~\cite{Lorenza98,Pokharel18,Jurcevic2021}. 
To reduce the effect of stochastic gate error in a circuit execution, we use the Mitiq package~\cite{LaRose21} to implement zero-noise extrapolation (ZNE) using random gate folding. 
This method scales the gate noise by performing gate folding $G\to GG^\dagger G$ on randomly chosen two-qubit gates throughout the circuit. 
This results in a noise scale factor $\lambda$ that can be noninteger, in contrast to the simpler global gate folding procedure described in Sec.~\ref{sec:Pulse-Level Implementation of the Ising Interaction}.
ZNE is best justified in the case where gate errors result in a stochastic quantum channel.
For this reason, we also implement Pauli twirling~\cite{Silva-Pauli_twirling-PRA-2008,Temme17, Li17}, in which two-qubit gates are dressed with random Pauli gates chosen so as not to affect the outcome of a circuit execution in the zero-noise limit. Averaging the results of many of these random circuit instances reduces the gate noise to a stochastic form.
The details of our implementation of these techniques are explained in more depth in Appendix~\ref{app:Mitigation}.

Finally, to enhance the signatures of the characteristic oscillatory dynamics, we implement postselection of measurement data to exclude CB measurement outcomes in which nearest-neighbor sites were measured to be in the state $\ket{11}$. 
As discussed in Sec.~\ref{sec:Model and Observables}, the probability of such configurations appearing when evolving the N\'eel state under the Hamiltonian \eqref{eq:H} is heavily suppressed when $V/\Omega$ is large. 
In Appendix~\ref{app:Mitigation}, we show strong numerical evidence that this is the case for the ratio $V/\Omega \approx 4.17$ used in this work. 
There we also show, however, that the Trotter error due to the large step size $\Delta t=1$ substantially increases the probability of generating these ``forbidden" configurations.
Thus, postselection mitigates the effect of both Trotter and gate error.
In this paper we will always compare QPU results using postselected data with exact results in which the Trotter dynamics generated by Eq.~\eqref{eq:Trotter} are projected into the Fibonacci Hilbert space before calculating the observable of interest.

We note that the set of error mitigation techniques we use for this work (including the scaled-$R_{ZX}$ implementation of the $R_{ZZ}$ gate) is similar to that used in Ref.~\cite{Kim21}. We highlight here a few important differences upon which we expand in Appendix~\ref{app:Mitigation}. First, Ref.~\cite{Kim21} takes a different approach to ZNE wherein the noise is scaled by scaling the duration and amplitude of the cross-resonance pulses. In contrast, our ZNE scheme treats the scaled-$R_{ZX}$ pulse schedule for the $R_{ZZ}(\theta)$ gate as a custom gate which is folded in the same way as other gates, including CNOT. Ref.~\cite{Kim21} also takes a different approach to Pauli twirling of the $R_{ZZ}(\theta)$ gate, which is a non-Clifford gate for generic $\theta$. In particular, Ref.~\cite{Kim21} performs Pauli twirling using only random Pauli operators from the set $\{II,XX,YY,ZZ\}$. Our Pauli twirling method, described in Appendix~\ref{app:Mitigation}, uses the full set of two-qubit Pauli operators to perform the twirling, and we prove that it results in a stochastic noise channel.

\section{Error-Mitigated Results}
\label{sec:Error-Mitigated Results}
\subsection{$Z_{\pi}$ and Loschmidt Echo}
\label{sec:Zpi and Losch Results}

\begin{figure}[t]
\includegraphics[width=1.0\columnwidth]{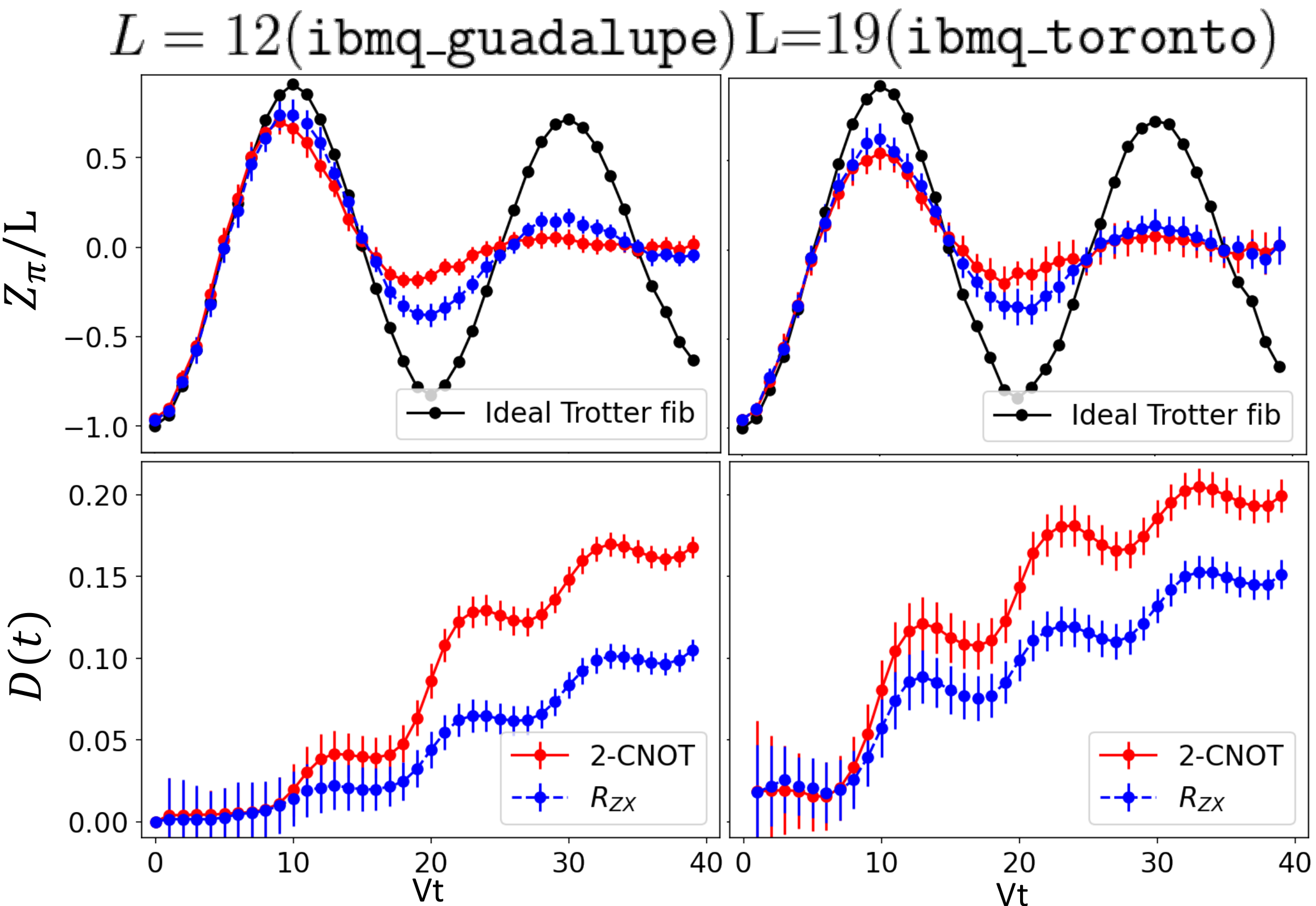}
\centering
\caption{
Error-mitigated Trotter simulation of the staggered magnetization density $\braket{Z_\pi(t)}/L$ [(a),(b); see Eq.~\eqref{eq:Zpi}] and accumulated error $D(t)$ [(c),(d); see Eq.~\eqref{eq:Dt}] from the initial state $\ket{Z_2}$. Data for chains of 12 [(a),(c)] and 19 [(b),(d)] qubits obtained using \texttt{ibmq\_guadalupe} and \texttt{ibmq\_toronto}, respectively. Data for both the two-CNOT (red) and scaled-$R_{ZX}$ (blue) implementations of the $R_{ZZ}$ gate are shown, with noiseless Fibonacci-projected Trotter simulation results (black) for reference.
Each data point in (a) and (b) is the result of linear ZNE with scale factors $\lambda\in\{1.0,1.5,2.0\}$ for 10 random Pauli-twirling circuits.
Error bars for all data points represent uncertainty in the ZNE and are calculated as described in the main text.
The error mitigation strategies outlined in Sec.~\ref{sec:Error Mitigation Techniques} result in a substantial improvement for both implementations of the $R_{ZZ}$ gate relative to the results shown in Fig.~\ref{fig:Z_pi_nomit}.
}
\label{fig:Z_pi_mit}
\end{figure}
 
 We now test the degree to which the error mitigation strategies outlined in Sec.~\ref{sec:Error Mitigation Techniques} improve the results shown in Fig.~\ref{fig:Z_pi_nomit}. The results of fully error-mitigated calculations of $\braket{Z_\pi(t)}/L$ and $D(t)$ on \texttt{ibmq\_guadalupe} and \texttt{ibmq\_toronto} are shown in Fig.~\ref{fig:Z_pi_mit}.
 For these simulations, we performed ZNE with random gate folding scale factors $\lambda\in\{1,1.5,2.0\}$ and perform Pauli twirling with 10 random circuit instances. As in Fig.~\ref{fig:Z_pi_nomit}, we evolve for 39 Trotter steps and use 8192 shots per circuit execution. Unlike Fig.~\ref{fig:Z_pi_nomit}, we use postselected QPU data and compare with the Fibonacci-projected Trotter evolution (black). With the extra circuits needed to perform ZNE and Pauli twirling, the total number of circuits run on each device is now $40\times 10\times 3=1200$ (including $t=0$). ZNE is performed with a linear extrapolation to $\lambda=0$ for each Trotter step, with 10 data points for each scale factor. Each data point in Figs.~\ref{fig:Z_pi_mit}(a) and (b) corresponds to the value of the y-intercept obtained from the extrapolation, and the error bars on each point are standard deviations calculated from the covariance matrix of the data set. In Figs.~\ref{fig:Z_pi_mit} (c) and (d), the error bars on $D(t)$ [see Eq.~\eqref{eq:Dt}] are standard deviations calculated by propagation of errors.
 In Appendix~\ref{app:Qubits}, we show data for individual qubits to illustrate how the simulation quality varies from qubit to qubit.
 
\begin{figure}[t!]
\includegraphics[width=.9\columnwidth]{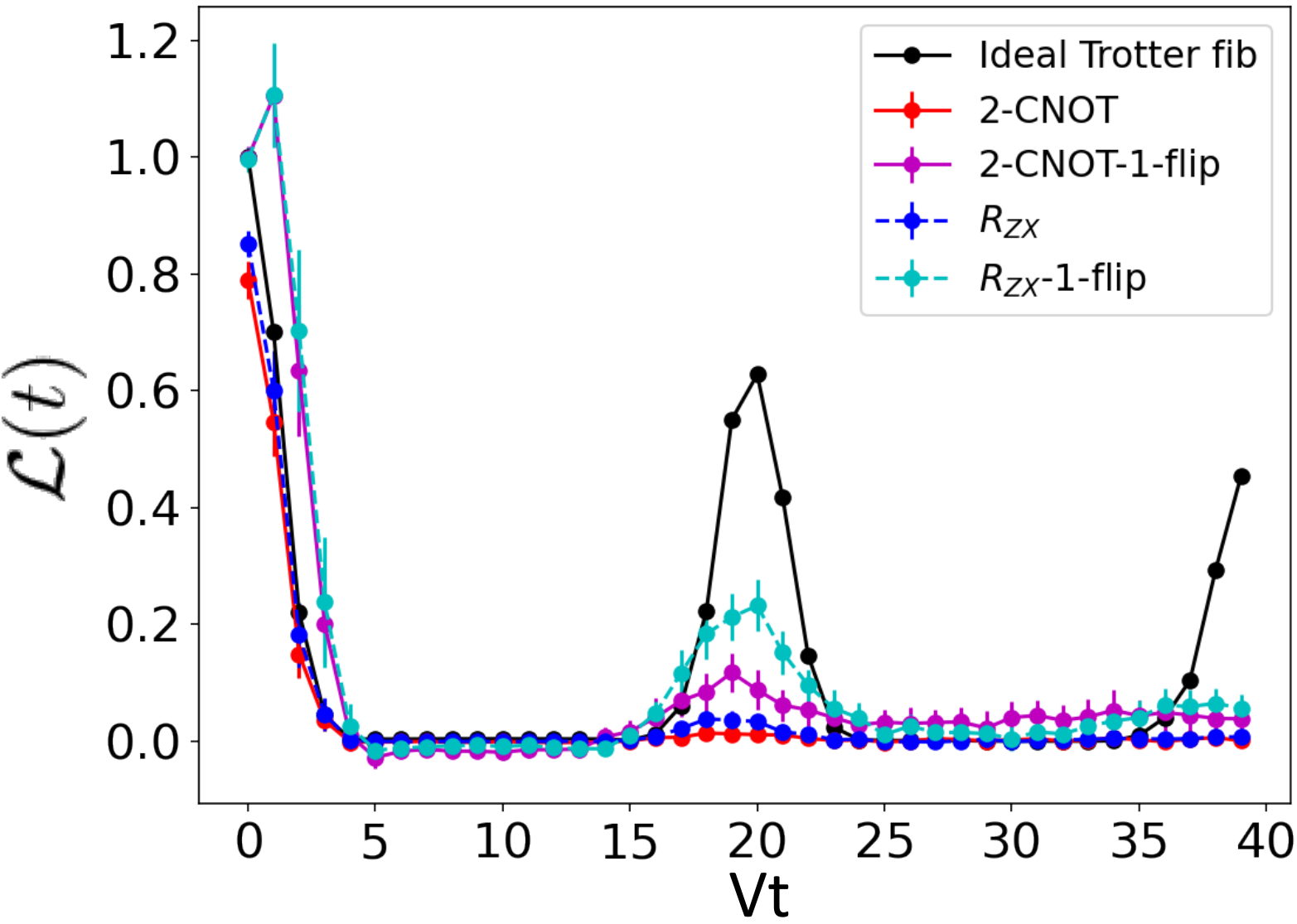}
\centering
\caption{
Loschmidt echo $\mathcal L(t)$ of the initial state $\ket{Z_2}$ for a chain of 12 qubits calculated on \texttt{ibmq\_guadalupe} using the same error mitigation techniques described in Fig.~\ref{fig:Z_pi_mit} and the text. Data obtained from simulations using the two-CNOT implementation of the $R_{ZZ}$ gate (red) barely show any tendency toward a revival around $Vt=20$, whereas data obtained using the scaled-$R_{ZX}$ implementation (blue) show a more pronounced revival. Both revivals are nowhere near as pronounced as the one obtained from the Fibonacci-projected ideal Trotter simulation (black), further indicating the effect of errors on the simulation despite the error-mitigation measures used.
}
\label{fig:losch}
\end{figure}
 
 Figs.~\ref{fig:Z_pi_mit} (a) and (b) show a substantial improvement over the results shown in Figs.~\ref{fig:Z_pi_nomit} (a) and (b). The scaled-$R_{ZX}$ implementation of $R_{ZZ}$ still outperforms the two-CNOT implementation. In particular, for the 12-qubit calculation performed on \texttt{ibmq\_guadalupe}, the scaled-$R_{ZX}$ approach yields a staggered magnetization density $\braket{Z_\pi(t)}/L$ that is in good agreement with the Trotter simulation until roughly $Vt=15$. This is roughly a twofold improvement relative to the case without error mitigation. For both the 12- and 19-qubit calculations, the scaled-$R_{ZX}$ approach yields visible oscillations up to the final time $Vt=39$, which covers about two oscillation periods. In contrast, even with error mitigation the two-CNOT implementation of the Trotter circuit yields visible oscillations only over one period. In Figs.~\ref{fig:Z_pi_mit} (c) and (d), we see that the accumulated error $D(t)$ is substantially reduced as compared to the unmitigated results shown in Fig.~\ref{fig:Z_pi_nomit}. For the scaled-$R_{ZX}$ implementation of the Trotter circuit, error mitigation leads to a reduction of the total accumulated error $D(t=39V^{-1})$ by $56\%$ ($29\%$) for the 12-qubit (19-qubit) calculation. In contrast, error mitigation of the two-CNOT implementation of the Trotter circuit results in a $37\%$ ($22\%$) error reduction for the 12-qubit (19-qubit) calculation.
 Compared to results where only postselection is applied (see Appendix~\ref{app:Mitigation}), the results in Fig.~\ref{fig:Z_pi_mit} show an error reduction of $37\%$ ($29\%$) for the scaled-$R_{ZX}$ implementation and $23\%$ ($20\%$) for the two-CNOT implementation for the 12-qubit (19-qubit) calculation.
 As in Fig.~\ref{fig:Z_pi_nomit} (c) and (d), the rate of error accumulation for the scaled-$R_{ZX}$ implementation is always less than it is for the two-CNOT implementation. Interestingly, with error mitigation the final accumulated error is larger for the 19-qubit calculation than for the 12-qubit calculation [cf. Fig.~\ref{fig:Z_pi_nomit}].

To further test the extent to which error mitigation improves the accuracy of the results obtained from the QPUs, we calculate the Loschmidt echo $\mathcal L(t)$ [see Eq.~\eqref{eq:L}] for a 12-qubit system using the same data set from which the results of Fig.~\ref{fig:Z_pi_mit} (a) and (c) were obtained. The results are shown in Fig.~\ref{fig:losch}. The scaled-$R_{ZX}$ implementation of the Trotter circuit shows a faint but noticeable revival near the first oscillation period $Vt= 2\pi/(1.33\Omega)\approx19.68$, while the two-CNOT implementation shows hardly any revival at all. Note that the fact that the Loschmidt echo is measured to take a finite value on the device after $\sim 20$ Trotter steps is remarkable given the exponential sensitivity of the Loschmidt echo to changes in the state $\ket{\psi(t)}$. The fact that the scaled-$R_{ZX}$ implementation of the Trotter circuit yields a substantially enhanced revival is further evidence of the performance advantage offered by that approach.

To enhance the Loschmidt echo signal, we also consider the effect of counting shots in which the measurement outcome differs from the N\'eel state $\ket{Z_2}$ by a single bit flip. This makes the metric $\mathcal L(t)$ more forgiving by counting instances where the system \emph{almost} returns to the initial state. The results, shown in Fig.~\ref{fig:losch}, demonstrate that this protocol indeed boosts the amplitude of the first revival in $\mathcal L(t)$. We observe greater enhancement of the first revival for the scaled-$R_{ZX}$ implementation of $R_{ZZ}$ than for the two-CNOT implementation, consistent with our other results. In both cases, the signal enhancement is localized in time near the first revival time but becomes more diffuse at later times.

\subsection{Connected Correlation Function}
\label{sec:Corr}
We now discuss how we measure the nontrivial correlation function $\mathcal C_Y(t)$ [see Eq.~\eqref{eq:CY}] on a QPU. The correlation function $\mathcal C_Y(t)$ is of the form $\braket{O(t)O(0)}$ where $O(t)=e^{iHt}O e^{-iHt}$ is a Hermitian operator that can be expanded in the basis of Pauli strings. One way to measure such a correlator on a quantum computer is to use a so-called indirect measurement technique based on the Hadamard test~\cite{Ortiz01,Somma02}. This approach uses an ancilla qubit and relies on the ability to apply Pauli gates controlled by the ancilla. For an operator like $O=Y_\pi$, which is a sum of many local Pauli strings, this means that the ancilla qubit must be able to couple to all qubits in the chain. Achieving this on superconducting qubit QPUs with nearest-neighbor connectivity requires substantial gate overhead, making this approach somewhat impractical for our purposes. We therefore opt instead for a direct measurement approach, which avoids the use of ancilla qubits at the cost of running more circuits with fewer gates. We now describe this approach, which is based on the proposal of Ref.~\cite{Mitarai19}.

\begin{figure}[t!]
\includegraphics[width=0.9\columnwidth]{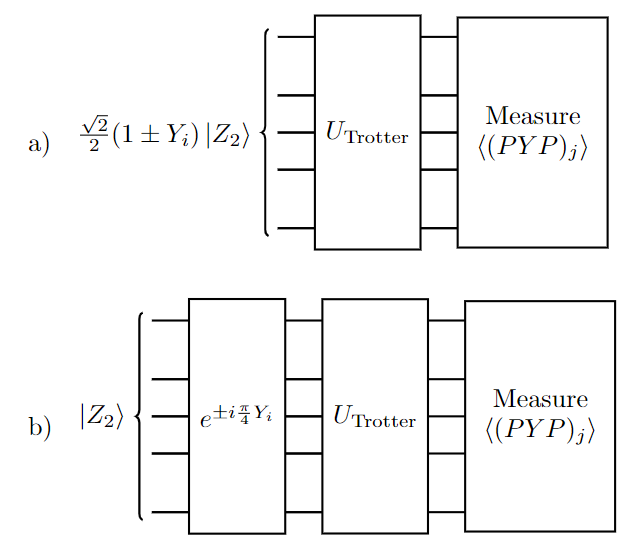}
\centering
\caption{
Circuits used to calculate the correlation function $\mathcal C_Y(t)$ on the QPU. 
(a) Circuit to calculate $\left\langle (PYP)_{i}\right\rangle_{M_{Y_{i}}=\pm1}$ [Eq.~\eqref{eq:MF2}].
(b) Circuit to calculate $\left\langle (PYP)_{i}\right\rangle_{\pm Y_{i}}$ [Eq.~\eqref{eq:MF3}]. Here, $U_{\text{Trotter}}$ denotes the Trotter circuit. In the beginning of each circuit, a different initial state is prepared. At the end of each circuit, the expectation value of $(PYP)_j$ is measured.
}
\label{fig:Cor_cirs}
\end{figure}

$\mathcal C_{Y}(t)$ can be calculated as a sum of many local correlators, i.e.,
\begin{equation}
    \mathcal C_Y(t)=\sum_{i,j}\left(-1\right)^{i+j}\bra{Z_{2}}(PYP)_j(t)(PYP)_i\ket{Z_{2}}\,,
\end{equation}
where the operators $(PYP)_i$ are defined in Eq.~\eqref{eq:PYP_def}.
This can be further simplified using our knowledge that the initial state is $\ket{Z_2}$, since $(PYP)_{i}\ket{\ldots(010)_i\ldots}=Y_{i}\ket{\ldots(010)_i\ldots}$ and $\left(PYP\right)_{i}\ket{\ldots(101)_i\ldots}=0$. Therefore, we can simplify the above to
\begin{align}
    \mathcal C_Y(t)=\sum_{j}\sum_{i\text{ even}}\left(-1\right)^{i+j}\bra{Z_{2}}(PYP)_j(t)Y_i\ket{Z_{2}}.
\end{align}
It remains to evaluate the local correlators $\bra{Z_{2}}(PYP)_{j}(t)Y_{i}\ket{Z_{2}}$. In Ref.~\cite{Mitarai19}, it is shown that these can be calculated as
\begin{widetext}
\begin{subequations}
\label{eq:MF}
\begin{align}
\label{eq:MF1}
\begin{split}
    &\bra{Z_{2}}(PYP)_{j}(t)Y_{i}(0)\ket{Z_{2}}\\ 
    &\qquad\qquad=
    \frac{1}{2}\left [\bigl\langle (PYP)_{j}(t)\bigr\rangle _{M_{Y_i}=1} -\bigl\langle (PYP)_{j}(t)\bigr\rangle _{M_{Y_i}=-1}\right] 
    -\frac{i}{2}\Bigl[\bigl\langle (PYP)_{j}(t)\bigr\rangle _{+Y_i}-\bigl\langle (PYP)_{j}(t)\bigr\rangle _{-Y_i}\Bigr],
\end{split}
\end{align}
with
\begin{align}
\label{eq:MF2}
    \bigl\langle (PYP)_{j}(t)\bigr\rangle _{M_{Y_{i}}=\pm1}
    =\frac{1}{2}\bra{Z_{2}}\left(\frac{I\pm Y_{i}}{2}\right)U^\dagger(t)(PYP)_{j}U(t)\left(\frac{I\pm Y_{i}}{2}\right)\ket{Z_{2}}
\end{align}
and 
\begin{align}
\label{eq:MF3}
    \left\langle (PYP)_{j}(t)\right\rangle _{\pm Y_i}
    =\bra{Z_{2}}e^{\mp i\frac{\pi}{4}Y_{i}}U^\dagger(t)(PYP)_{j}U(t)e^{\pm i\frac{\pi}{4}Y_{i}}\ket{Z_{2}}\,.
\end{align}
\end{subequations}
 \end{widetext}
 In the above expressions, $U(t)$ is the evolution operator out to time $t$, which is approximated on the QPU by the Trotter circuit. Note that both Eqs.~\eqref{eq:MF2} and \eqref{eq:MF3} can be formulated as the expectation value of $(PYP)_j$ in a particular time-evolved state. Quantum circuits to evaluate these expectation values are shown in Fig.~\ref{fig:Cor_cirs}. To prepare the initial state $\frac{\sqrt{2}}{2}(I\pm Y_i)\ket{Z_2}$ needed to evaluate Eq.~\eqref{eq:MF2} on the device, we act on the $i$th qubit with either an identity or an $X$ gate (depending on the choice of $+$ or $-$, respectively), followed by a Hadamard gate and an $S$ gate. (Note that $i$ is even, so the initial state of the $i$th qubit is always $\ket{1}$ in the $\ket{Z_2}$ state).
 
 We have calculated $\mathcal C_Y(t)$ for chains of $L=5$ and $L=12$ sites on \texttt{ibmq\_casablanca} and \texttt{ibmq\_guadalupe}, respectively, for both the QMBS regime ($V=\Delta t=1$ and $\Omega=0.24$) and the chaotic regime, where we use parameters $V=1$, $\Omega=2$ and $\Delta t = 0.16$. The calculation of $\mathcal C_Y(t)$ on the QPU proceeds as follows. For each $i$ and $j$, we need to evaluate the four circuits shown in Fig.~\ref{fig:Cor_cirs} to calculate $\bigl\langle (PYP)_{j}(t)\bigr\rangle_{M_{Y_{i}}=\pm1}$ and $\left\langle (PYP)_{j}(t)\right\rangle_{\pm Y_i}$. Note that $[(PYP)_j,(PYP)_{j+2}]=0$, so we can measure $(PYP)_j$ for all even and all odd $j$ in one shot---in fact, this is one of the advantages of the direct measurement scheme. The total number of circuits needed to calculate $\mathcal C_Y(t)$ is therefore $4\times 2\times\lfloor L/2\rfloor$, where $\lfloor \cdot\rfloor$ denotes the floor function. For each of these circuits, we employ ZNE with random gate folding for scale factors $\lambda\in\{1.0,1.5,2.0\}$. For the $L=5$ calculations, we employ Pauli twirling with 8 random circuit instances in both the QMBS and chaotic regimes for each scale factor used in ZNE. For the $L=12$ calculations, we used 10 random circuit instances for the QMBS regime and 9 for the chaotic regime. For all cases, we evolve the system for $30$ Trotter steps and measure the system with $8192$ shots at each step~\footnote{In this case, since we are not measuring in the $Z$-basis, we do not perform postselection on the QPU data. We therefore compare with the ideal Trotter dynamics rather than the projected dynamics.}. For the $5$-qubit system, we evaluate a total of $11520$ circuits to measure $\mathcal C_Y(t)$; for the $12$-qubit system, we evaluate $43200$ circuits for the QMBS case and $38880$ circuits for the chaotic case. We use the scaled-$R_{ZX}$ implementation of the Trotter circuit, since the results of Sec.~\ref{sec:Zpi and Losch Results} indicate that this approach outperforms the two-CNOT implementation.
 
 \begin{figure}[t]
\includegraphics[width=1.0\columnwidth]{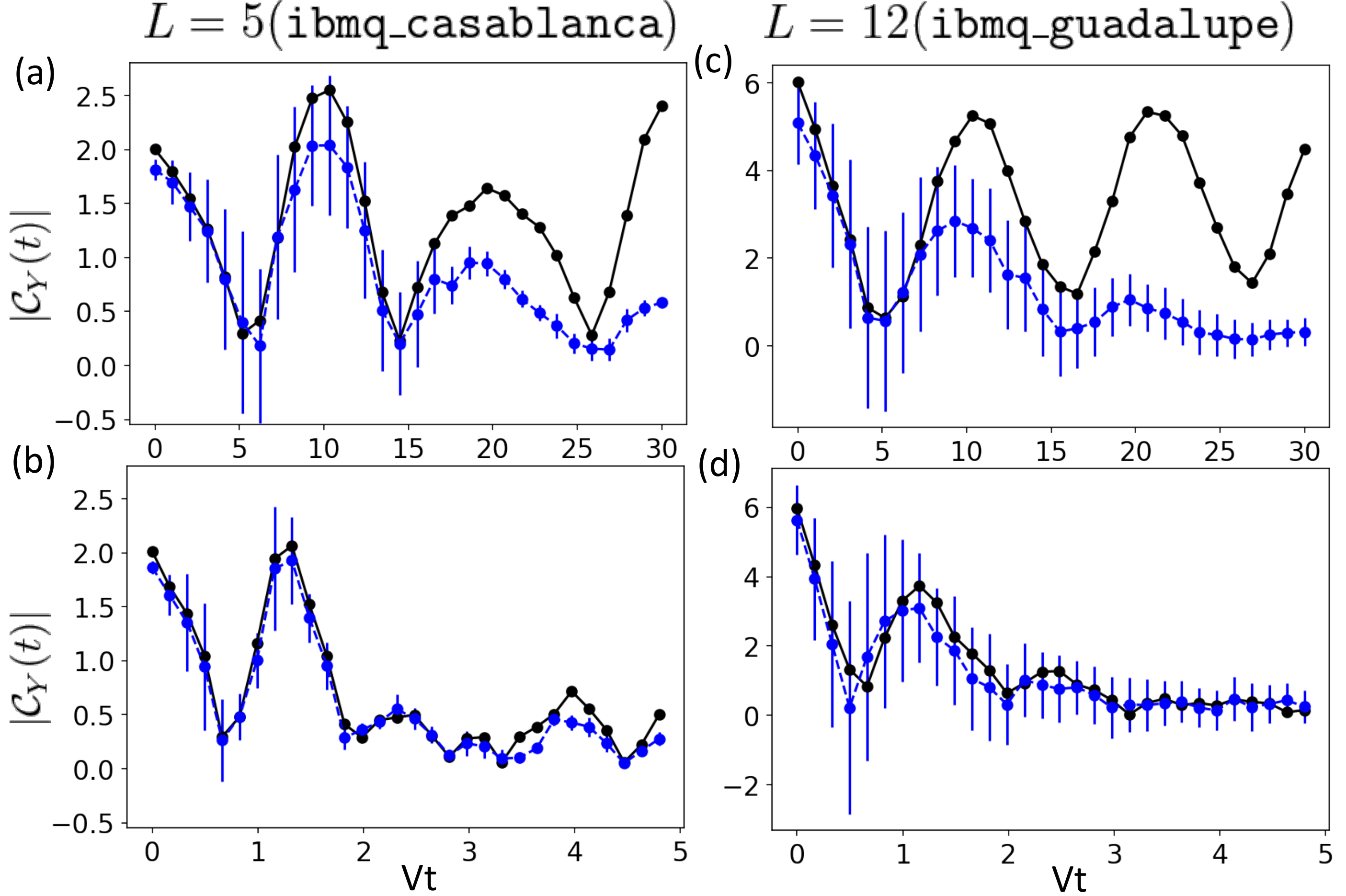}
\centering
\caption{
Dynamics of the correlator $\mathcal C_Y(t)$ [see Eq.~\eqref{eq:CY}] from the initial state $\ket{Z_2}$ in (a,c) the QMBS regime ($V=\Delta t=1$, $\Omega=0.24$) and (b,d) the chaotic regime ($V=1$, $\Omega=2$, $\Delta t=0.16$). Panels (a,b) are calculated for a chain of 5 qubits using \texttt{ibmq\_casablanca}, and panels (c,d) are calculated for a chain of 12 qubits using \texttt{ibmq\_guadalupe}. The calculation uses ZNE for scale factors $\lambda\in\{1.0,1.5,2.0\}$ with 8 (a,b), 10 (c), and 9 (d) random circuit instances for Pauli twirling. Error bars representing the uncertainty in the ZNE were calculated as in Fig.~\ref{fig:Z_pi_mit}. For $L=5$, oscillations with relatively slowly decaying amplitude are clearly visible throughout the simulation time window in the QMBS regime (a). For $L=12$, these oscillations remain coherent but exhibit a more rapid decay due to the accumulation of gate and readout errors. In the chaotic regime (b,d), the correlator rapidly decays after a single approximate revival and exhibits good agreement with the ideal Trotter simulation results over the full simulation time window.
}
\label{fig:Cor}
\end{figure}
 
 The results of these calculations are shown in Fig.~\ref{fig:Cor}(a,c) for the QMBS regime and in panels (b,d) for the chaotic regime. We show results for $|\mathcal C_Y(t)|$ for ease of visualization, as Eqs.~\eqref{eq:MF} demonstrate that $\mathcal C_Y(t)$ is generically complex. [Error bars on individual data points are calculated as described at the beginning of Sec.~\ref{sec:Zpi and Losch Results}.] In Fig.~\ref{fig:Cor}(a), we see that for $L=5$ the calculation of $\mathcal C_Y(t)$ exhibits good quantitative agreement with the ideal Trotter calculation out to $Vt\approx 15$, and qualitative agreement is maintained throughout the full time window until $Vt=30$. In particular, oscillations with the expected period $\pi/(1.33\Omega)\approx 9.84$ are visible throughout the time evolution window (recall that we plot the absolute value of the correlation function). In contrast, in the chaotic regime for $L=5$ shown in Fig.~\ref{fig:Cor}(b), we find that $|\mathcal C_Y(t)|$ exhibits one approximate revival followed by a rapid decay and incoherent dynamics after time $Vt=2$. Note that, in the chaotic regime, Trotter dynamics involves a smaller $R_{ZZ}$ rotation angle than in the QMBS regime, resulting in shorter cross resonance pulse durations and higher $R_{ZZ}$ gate fidelities. Consequently, the dynamics exhibit excellent quantitative agreement with the ideal Trotter simulation for approximately 22 Trotter steps.
 The results of the $L=12$ calculation are shown in Fig.~\ref{fig:Cor}(c) and (d) for the QMBS and chaotic regimes, respectively. While the results for the chaotic regime remain in very good agreement with the ideal Trotter simulation due to the smaller $R_{ZZ}$ rotation angle discussed above, the results in the QMBS regime begin to differ from the ideal Trotter results around $Vt=9$. This is likely due to the fact that the calculation of $\mathcal C_Y(t)$ involves summing $O(L^2)$ terms, each of which suffers from gate and readout errors and is the result of a separate zero noise extrapolation. Despite the more drastic accumulation of error for $L=12$, underdamped oscillations close to the correct frequency are clearly visible throughout the simulation time window.
 These results provide a clear demonstration that the coherence \textit{and} long-range many-body correlations present in the dynamics of the N\'eel state in the QMBS regime can be probed on current quantum devices.

\section{Conclusions and Outlook}
\label{sec:Conclusions and Outlook}

In this work, we have used pulse-level control and a variety of quantum error mitigation techniques to simulate the dynamics of a spin chain with QMBS on IBM QPUs for chains of up to 19 qubits. QMBS constitute an intriguing quantum dynamical regime characterized by nontrivial many-body coherence and long-range correlations. We probed this physics by measuring the dynamics of three quantities: the staggered magnetization $\braket{Z_\pi(t)}$, the Loschmidt echo $\mathcal L(t)$, and the connected unequal-time correlation function $C_Y(t)$. We found that $\braket{Z_\pi(t)}$ and $\mathcal C_Y(t)$ exhibit reasonable quantitative agreement with the ideal Trotter simulation at early times, and visible oscillations with the correct frequency over $39$ Trotter steps. In contrast, the Loschmidt echo $\mathcal L(t)$ exhibits only the faintest of revivals, indicating the substantial impact of various noise sources including thermal relaxation and gate and readout errors. Nevertheless, the qualitative features of QMBS are pronounced on time scales beyond which $\mathcal L(t)$ decays to zero, indicating the presence of coherent many-body dynamics with long-range correlations that cannot be explained by the precession of free spins. To obtain these results, we found it essential to use a pulse-level implementation of the Trotterized Ising interaction relying on amplitude- and duration-scaled cross-resonance pulses and to apply a number of error mitigation techniques. 

These results provide a physics-based benchmark of QPU performance in the NISQ era. Thus, it would be interesting to repeat these experiments on systems with higher quantum volume ($QV$), which is a metric that takes into account the number of qubits as well as the error rate of the device~\cite{Moll18}. Such devices include \texttt{ibmq\_washington} (127 qubits, $QV=64$) or \texttt{ibmq\_kolkata} (27 qubits, $QV=128$), which was used in Ref.~\cite{Kim21}. In contrast, the IBM QPUs used for our 12- and 19-qubit simulations, \texttt{ibmq\_guadalupe} and \texttt{ibmq\_toronto}, have $QV=32$. Furthermore, given the variety of available tools for error mitigation, it will be important to undertake a systematic exploration of the optimal implementations of techniques like ZNE and Pauli twirling for non-Clifford gates defined at the pulse level. As part of this effort, it would also be worthwhile to consider alternatives to ZNE including probabilistic error cancellation~\cite{Temme17,Zhang20,Sun21,caiMultiexponentialErrorExtrapolation2021, mariExtendingQuantumProbabilistic2021}, virtual distillation~\cite{hugginsVirtualDistillationQuantum2021,koczorExponentialErrorSuppression2021, Vovrosh-PRE-2021}, or Clifford data regression~\cite{Czarnik21,Sopena21}.

As devices with higher $QV$ become available, it will be interesting to pursue further the calculation of nontrivial multi-time correlation functions on quantum devices. One quantity that can be computed using the methods proposed in Ref.~\cite{Mitarai19} and further developed here is the transport of conserved quantities. For example, in the MFIM \eqref{eq:H}, the most natural conserved quantity to consider is the energy. The two-point correlation function of the energy density can be used to measure the timescales associated with energy transport. In the chaotic regime of the model, such transport is expected to be diffusive. Higher-order correlation functions associated with transport beyond the linear-response regime can also be considered, as well as out-of-time-ordered correlation functions (OTOCs) which can be used to characterize quantum chaos~\cite{Maldacena16,Huang16,Roberts17,Swingle17,Rozenbaum17}.

\acknowledgments
The authors acknowledge valuable discussions with M.~S.~Alam and N.~F.~Berthusen. This material is based upon work supported by the National Science Foundation under Grant No.~DMR-2038010. Calculations for spin models with more than seven sites on quantum hardware, and part of the associated analyses by Y.~Yao, are supported by the U.S. Department of Energy, Office of Science, National Quantum Information Science Research Centers, Co-design Center for Quantum Advantage (C2QA) under contract number DE-SC0012704. We acknowledge use of the IBM Quantum Experience, through the IBM Quantum Researchers Program. The views expressed are those of the authors, and do not reflect the official policy or position of IBM or the IBM Quantum team.

\begin{appendix}
\section{Details on Simulation Methods}
\label{app:Mitigation}

\subsection{Scaled-$R_{ZX}$ Implementation of the $R_{ZZ}$ Gate}

\begin{figure}[t]
\includegraphics[width=1.0\columnwidth]{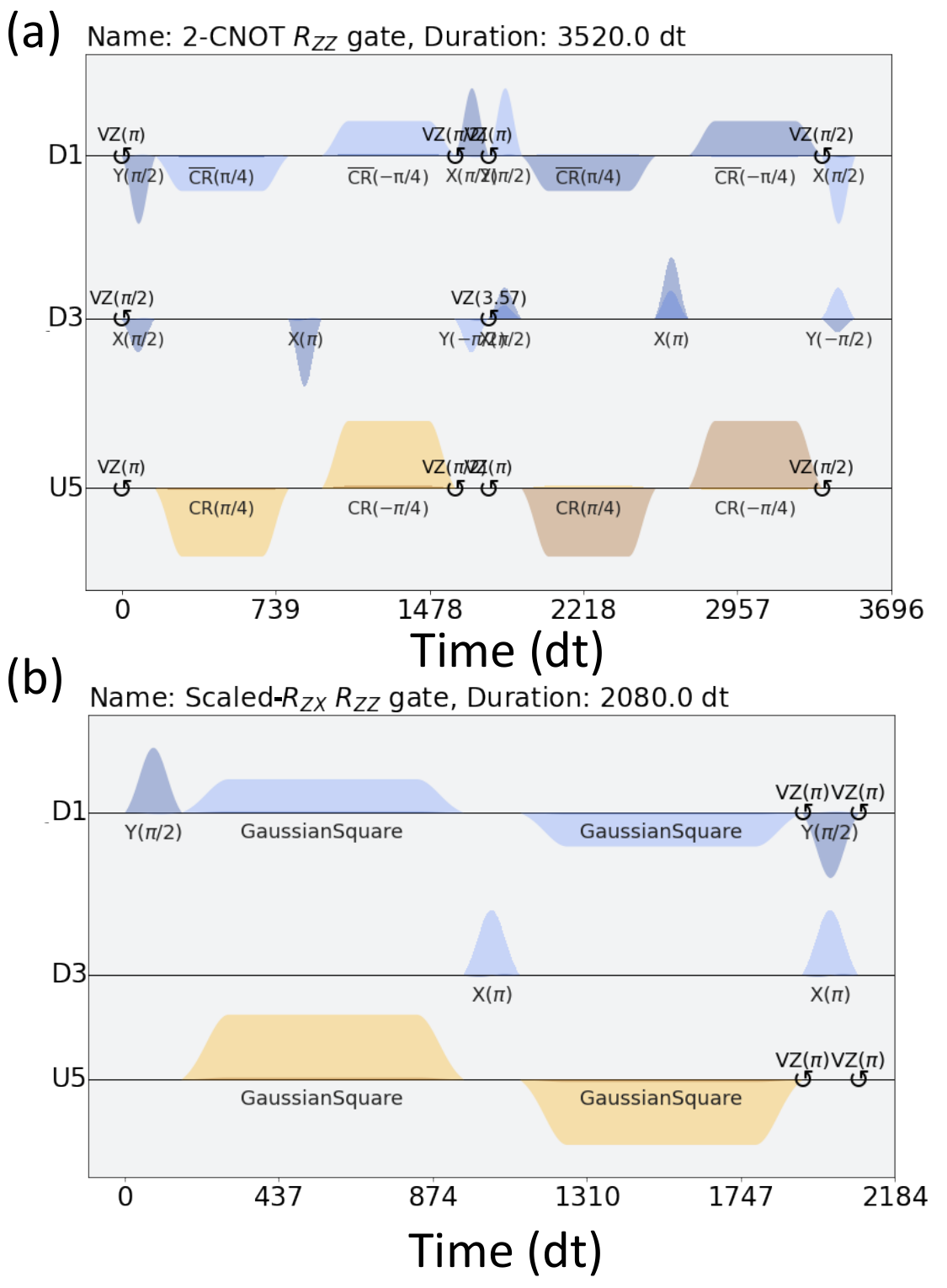}
\centering
\caption{
Pulse schedules for the two implementations of the $R_{ZZ}$ gate.
(a) Pulse schedule for the two-CNOT implementation of $R_{ZZ}(\theta=2.0)$ on \texttt{ibmq\_casablanca}. Pulse durations are measured in units of $dt=0.2222\text{ ns}$. Pulses labeled ``$\text{CR}(\pi/4)$" and ``$\overline{\text{CR}}(-\pi/4)$"  are the cross-resonance pulses described in the text. Other Gaussian pulses correspond to single-qubit gates. On the D1 channel, the overlapping symbols above two circular arrows read $VZ(\pi)$. This denotes that a ``virtual" rotation by $\pi$ around $Z$~\cite{McKay17} has been implemented using a pulse delay. The symbols below the two central Gaussian pulses read $X(\pi/2)$ (dark pulse) and $Y(\pi/2)$ (light pulse), respectively, indicating rotations by $\pi/2$ around the $X$ and $Y$ axes, respectively. On the D3 channel, the overlapping symbols below the two central Gaussian pulses read $Y(-\pi/2)$ (light pulse) and $X(\pi/2)$ (dark pulse), respectively.
(b) Pulse schedule for the scaled-$R_{ZX}$ implementation of $R_{ZZ}(\theta=2.0)$ on \texttt{ibmq\_casablanca}. Pulses labeled ``GaussianSquare" are the scaled cross-resonance pulses discussed in the text. The scaled-$R_{ZX}$ implementation uses half as many cross-resonance pulses as the two-CNOT implementation.
}
\label{fig:Schedule}
\end{figure}

In this section, we discuss how to scale the amplitude and duration of the cross resonance (CR) pulses used to realize the $R_{ZX}(\pi/2)$ to obtain the more general gate $R_{ZX}(\theta)$ with arbitrary rotation angle.
The $R_{ZX}(\pi/2)$ gates used to generate the CNOT gate are realized by echoed cross-resonance pulses composed of CR$(\pm \pi/4)$ sandwiching an X-echoed $\pi$ pulse on the control qubit to eliminate the $ZI$ and $IX$ terms in the cross-resonance Hamiltonian~\cite{Alexander20}. Rotary pulses are applied to the target qubit to suppress the interaction $IY$ and $ZZ$ terms~\cite{Sundaresan20}. The CR pulse has a square-Gaussian shape consisting of a square pulse with width $W(\frac{\pi}{2})$ whose boundaries are smoothed into Gaussians with standard deviation $\sigma$. We denote the amplitude of the pulse by $A(\frac{\pi}{2})$.
The total area enclosed by the square-Gaussian pulse is therefore
\begin{align}
    \alpha=\left|A\left(\frac{\pi}{2}\right)\right|W\left(\frac{\pi}{2}\right)+\left|A\left(\frac{\pi}{2}\right)\right|\sigma\sqrt{2\pi}\,\text{erf}\left(n_{\sigma}\right)
\end{align}
where $n_{\sigma}$ is the number of standard deviations of the Gaussian tails that are chosen to be included in the pulse shape.
An example of a pulse schedule for the two-CNOT implementation of the $R_{ZZ}$ gate on \texttt{ibmq\_casablanca} is shown in Fig.~\ref{fig:Schedule}(a). This pulse schedule contains one copy of the echoed CR pulse schedule described above for each CNOT gate.

To change the angle of rotation of the $R_{ZX}$ gate from $\pi/2$ to an arbitrary $\theta$, we follow the method outlined in Ref.~\cite{Stenger2021} (see also Ref.~\cite{Earnest21}). When $\theta>\frac{\pi}{2\alpha}|A(\frac{\pi}{2})|\sigma\sqrt{2\pi}\text{erf}(n_{\sigma})$, one can change the area under the square-Gaussian pulse to $\alpha(\theta)=\frac{\theta}{\pi/2}\alpha$ by adjusting the width of the square pulse as follows:
\begin{align}
    W(\theta)=\frac{2\alpha\theta}{\pi|A(\frac{\pi}{2})|}-\sigma\sqrt{2\pi}\text{erf}(n_{\sigma}).
\end{align}
If $\theta<\frac{\pi}{2\alpha}|A(\frac{\pi}{2})|\sigma\sqrt{2\pi}\text{erf}(n_{\sigma})$, one can set $W(\theta)=0$ and reduce the amplitude to
 \begin{align}
 |A(\theta)|=\frac{2\alpha\theta}{\pi\sigma\sqrt{2\pi}\text{erf}(n_{\sigma})}.
 \end{align}
 Therefore, just like in the circuit depicted in Fig.~\ref{fig:Rzz}(b), we can sandwich the $R_{ZX} (\theta)$ with $R_Y(\pi/2)$ and $R_Y(-\pi/2)$ pulses on the second channel to create the $R_{ZZ}(\theta)$ gate pulse schedule shown in Fig.~\ref{fig:Schedule}(b). As discussed in the main text, this pulse schedule can have a substantially shorter duration than the pulse schedule that implements $R_{ZZ}(\theta)$ using two CNOT gates.
 
 \subsection{Trotter Evolution}
 
 \begin{figure}[t]
\includegraphics[width=1.0\columnwidth]{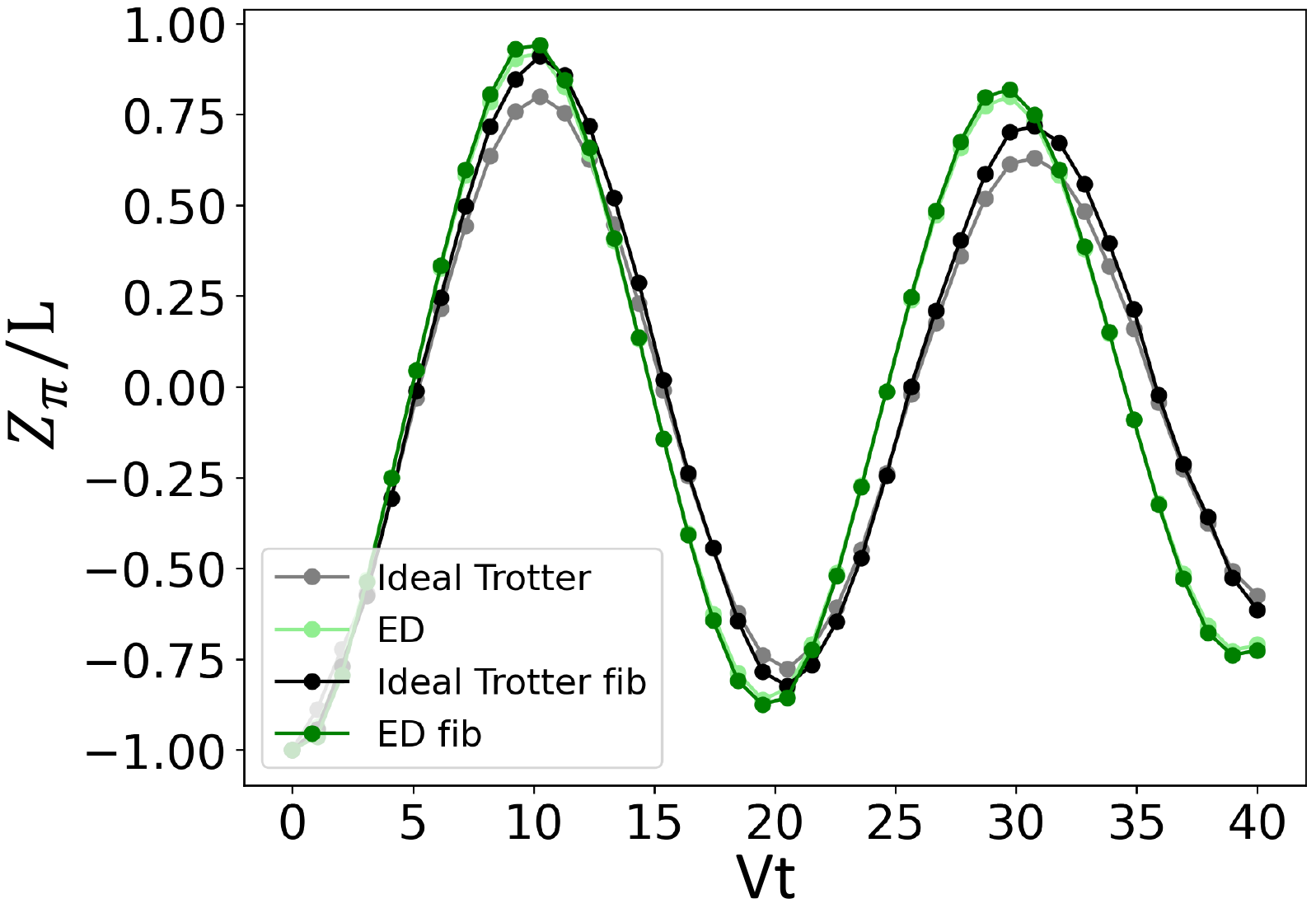}
\centering
\caption{Comparison of the dynamics of the staggered magnetization density $\braket{Z_\pi(t)}/L$ in the QMBS regime ($V=1$, $\Omega=0.24$) obtained from exact diagonalization (ED) (light green) and ideal Trotter simulations (grey) with $\Delta t=1$. Although the large time step incurs substantial Trotter error, clear long-lived oscillations of the staggered magnetization are visible for both evolutions. The green and black curves indicate Fibonacci-projected ED and ideal Trotter dynamics, respectively.
}
\label{fig:Z_pi_exact}
\end{figure}

In Fig.~\ref{fig:Z_pi_exact}, we compare exact diagonalization (ED) results for the dynamics of the staggered magnetization under the Hamiltonian \eqref{eq:H} with the results of a Trotter simulation at $L=12$ and $\Delta t=1$. Both simulations use model parameters $V=1$ and $\Omega=0.24$, which are appropriate for the QMBS regime. Significant differences between the ED and Trotter curves are visible starting around $Vt=5$. However, note that the Trotter circuit dynamics retains the coherent oscillations visible in the ED result. Thus, despite the large time step and the appreciable Trotter error, the Trotter circuit still exhibits strong signatures of QMBS.

We also plot for reference the dynamics of $\braket{Z_\pi}$ calculated with respect to the Fibonacci-projected ED and Trotter dynamics. The projection has much less effect on the ED dynamics than on the Trotter dynamics. This indicates that postselection of measurement outcomes (as described below) has a larger effect when a larger Trotter step size is used.

\subsection{Readout Error Mitigation}
To mitigate the effect of readout errors on QPU results, one can use readout error mitigation methods as described in Ref.~\cite{Bravyi21} and built into Qiskit Ignis. Suppose that $C_{\text{ideal}}$ is a vector containing the list of measurement counts for each computational basis state in the absence of readout error, and that $C_{\text{noisy}}$ is the same quantity with readout error.
The relationship between $C_{\rm ideal}$ and $C_{\rm noisy}$ can be characterized by a readout error matrix $M$ defined as
\begin{align} 
MC_{\text{ideal}}=C_{\text{noisy}}.
\end{align}
To obtain the ideal result from the noisy result, one can invert the readout error matrix: 
 \begin{align} 
C_{\text{ideal}}=M^{-1}C_{\text{noisy}}.
\end{align}
Qiskit Ignis supports several methods to obtain the readout error matrix $M$. One is to approximate $M$ as the tensor product of readout error matrices for each qubit as follows:
 \begin{align} 
M=\left[\begin{array}{cc}
1-\epsilon_{1} & \eta_{1}\\
\epsilon_{1} & 1-\eta_{1}
\end{array}\right]\otimes\cdots\otimes\left[\begin{array}{cc}
1-\epsilon_{n} & \eta_{n}\\
\epsilon_{n} & 1-\eta_{n}
\end{array}\right],
\end{align}
where $\epsilon_j$ and $\eta_j$ are the readout error rates for $0\rightarrow1$ and $1\rightarrow0$ respectively. This method is attractive because the $\epsilon_j$ and $\eta_j$ can be estimated by executing two circuits to prepare the states $\ket{0\dots 0}$ and $\ket{1\dots 1}$ and then measuring all qubits in the CB. Another method that we call ``complete readout error mitigation" prepares and measure all $2^L$ CB states from $\ket{0\dots 0}$ to $\ket{1\dots 1}$. This allows the elementwise extraction of $M$ for all computational basis states, but is more costly to perform as it involves measuring exponentially many CB states.  In this paper, we apply complete readout error mitigation for smaller systems ($L=5$) and use the tensor-product approximation for larger systems ($L=12, 19$). 

\subsection{Postselection}
 \begin{figure*}[t]
\includegraphics[width=1.0\textwidth]{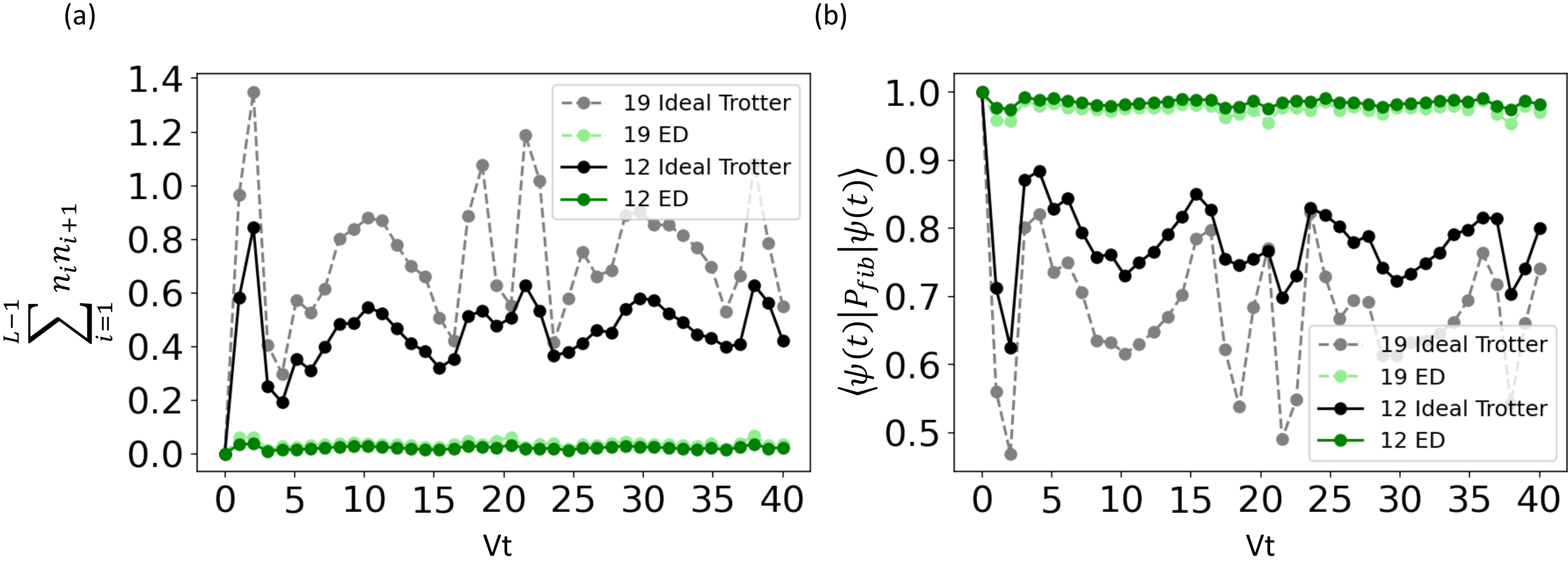}
\centering
\caption{
(a) The dynamics of the number of nearest-neighbor pairs of Rydberg excitations, $\sum^{L-1}_{i=1}n_{i}n_{i+1}$, obtained from ED and ideal Trotter with $\Delta t = 1$ at $L=12$ and $L=19$. The number of such nearest-neighbor pairs is subextensive even for the Trotter dynamics. (b) The weight of the time-evolved state $\ket{\psi(t)}$ in the Fibonacci Hilbert space, $\braket{\psi(t)|P_{\rm fib}|\psi(t)}$, plotted using ED and ideal Trotter at $L=12$ and $L=19$. Even though the Trotter dynamics loses more weight in the subspace over the simulation window than the ED dynamics, the Trotter-evolved state retains a finite weight in the subspace at these system sizes. Taken together, these results justify our use of the postselection technique described in this Appendix.}
\label{fig:NN}
\end{figure*}

\begin{figure}[t]
\includegraphics[width=1.0\columnwidth]{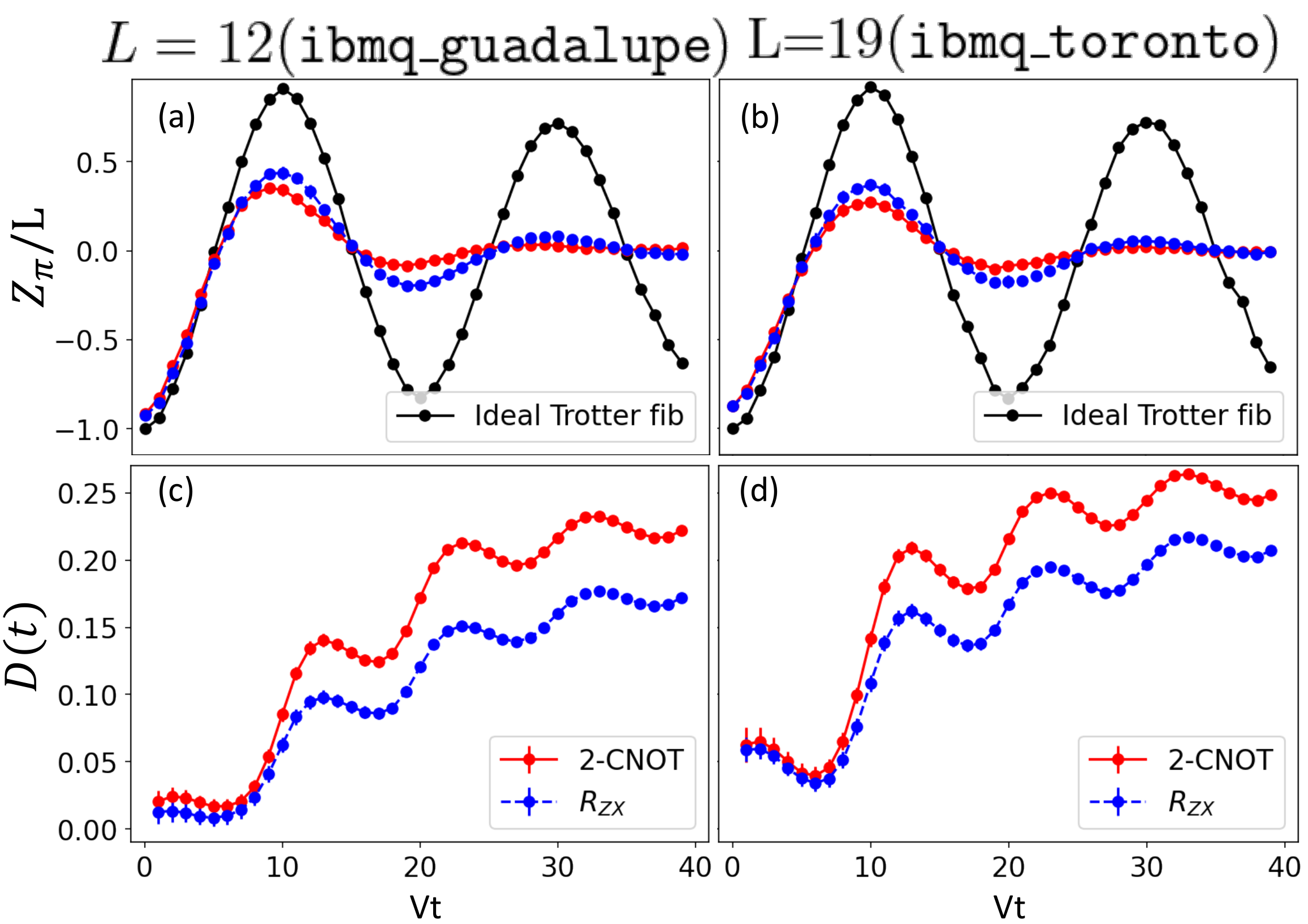}
\centering
\caption{
Postselected QPU results without further error mitigation. See the caption of Fig.~\ref{fig:Z_pi_nomit} for a description of panels (a)--(d). The red and blue curves are calculated using the same QPU dataset as Fig.~\ref{fig:Z_pi_nomit} with postselection applied, while the black curve denotes the Fibonacci-projected ideal Trotter dynamics for comparison. The data points on the black curve are used to calculate $D(t)$.  
}
\label{fig:Z_pi_nomit_pos}
\end{figure}

In the QMBS regime of the Hamiltonian \eqref{eq:H}, the probability of the system being in a CB state with two consecutive $1$s is heavily suppressed by the strong nearest-neighbor interaction, as discussed in the main text. Fig.~\ref{fig:NN}(a) shows the dynamics of $\braket{\sum^{L-1}_{i=1}n_{i}n_{i+1}}$ starting from the $\ket{Z_2}$ state under the Hamiltonian \eqref{eq:H} with $\Omega = 0.24 V$ for $L=12$ and $19$ (dark and light green curves, respectively). This indicates that the Hamiltonian dynamics generated by Eq.~\eqref{eq:H} produces a negligible number of pairs of consecutive $1$s over the times we aim to simulate with the quantum device. However, the large Trotter step $\Delta t=1$ used in our QPU simulations means that Trotter error can induce a more substantial growth of $\braket{\sum^{L-1}_{i=1}n_{i}n_{i+1}}$ starting from the same state, as is visible in the ideal Trotter dynamics curves for $L=12$ and $19$ (black and grey, respectively). Nevertheless, for both system sizes considered, the excitation-pair number $\braket{\sum^{L-1}_{i=1}n_{i}n_{i+1}}$ remains $\lesssim1$ over the course of the dynamics.

That $\braket{\sum^{L-1}_{i=1}n_{i}n_{i+1}}$ remains subextensive indicates that the Trotter dynamics retains a finite weight in the Fibonacci Hilbert space. This is confirmed by Fig.~\ref{fig:NN}(b), which plots the time-dependence of the weight of the time-evolved $\ket{Z_2}$ state in the Fibonacci space, $\braket{\psi(t)|P_{\rm fib}|\psi(t)}$. The Hamiltonian dynamics for $L=12$ and $19$ remain almost entirely in the restricted Hilbert space, while the Trotter dynamics with timestep $\Delta t = 1$ remain roughly $70$-$80\%$ within the space on average. 

Since the scarred dynamics we are trying to model take place primarily within the Fibonacci Hilbert space~\cite{Turner18a,Turner18b}, we take this as evidence that we can amplify their signatures by calculating observables with measurement outcomes postselected to lie within this space. To calculate the postselected dynamics, we remove CB measurement results containing two or more consecutive $1$s from the counting dictionary and calculate the expectation value using the rest of the data. 

The effect of postselection on our QPU results is shown in Fig.~\ref{fig:Z_pi_nomit_pos}, which plots the same dataset as in Fig.~\ref{fig:Z_pi_nomit} with postselection applied. These results already show an enhancement of the oscillatory signal relative to Fig.~\ref{fig:Z_pi_nomit}, even with no error mitigation measures besides postselection applied. The results of Fig.~\ref{fig:Z_pi_mit} clearly demonstrate the utility of applying further error mitigation techniques beyond postselection.

\subsection{Zero Noise Extrapolation and Pauli Twirling}
In order to reduce the effect of gate noise on the calculation of, e.g., an expectation value on a QPU, one can measure this expectation value at different noise scales and extrapolate to the zero-noise limit to estimate the ideal expectation value~\cite{Li17,Temme17,Kandala19, Giurgica-Tiron-ZNE-2020}. To increase the noise scale we can use unitary folding, which acts on a gate $G$ as 
\begin{align} 
G\mapsto GG^{\dagger}G.
\end{align}
Here $G^{\dagger}=G^{-1}$ so the above operation increases the depth of the circuit without changing its logical action on the input state in the unrealistic case where $G$ is implemented noiselessly on the QPU. In the realistic case where $G$ is noisy, this folding operation increases the effect of noise for that gate by a factor of $\sim 3$. In this paper, we use the Mitiq package~\cite{LaRose21} to randomly fold the gates in our circuit to achieve noise scale factors $\lambda\in\{1.0,1.5,2.0\}$ for the full circuit. Since Mitiq does not support folding of the $R_{ZZ}(\theta)$ gate, we first generate folded circuits using CNOT gates as placeholders for $R_{ZZ}(\theta)$ gates, and then replace all CNOT gates with $R_{ZZ}(\theta)$ and $R_{ZZ}(-\theta)$ gates as appropriate. 

\begin{figure}[t!]
\includegraphics[width=1.0\columnwidth]{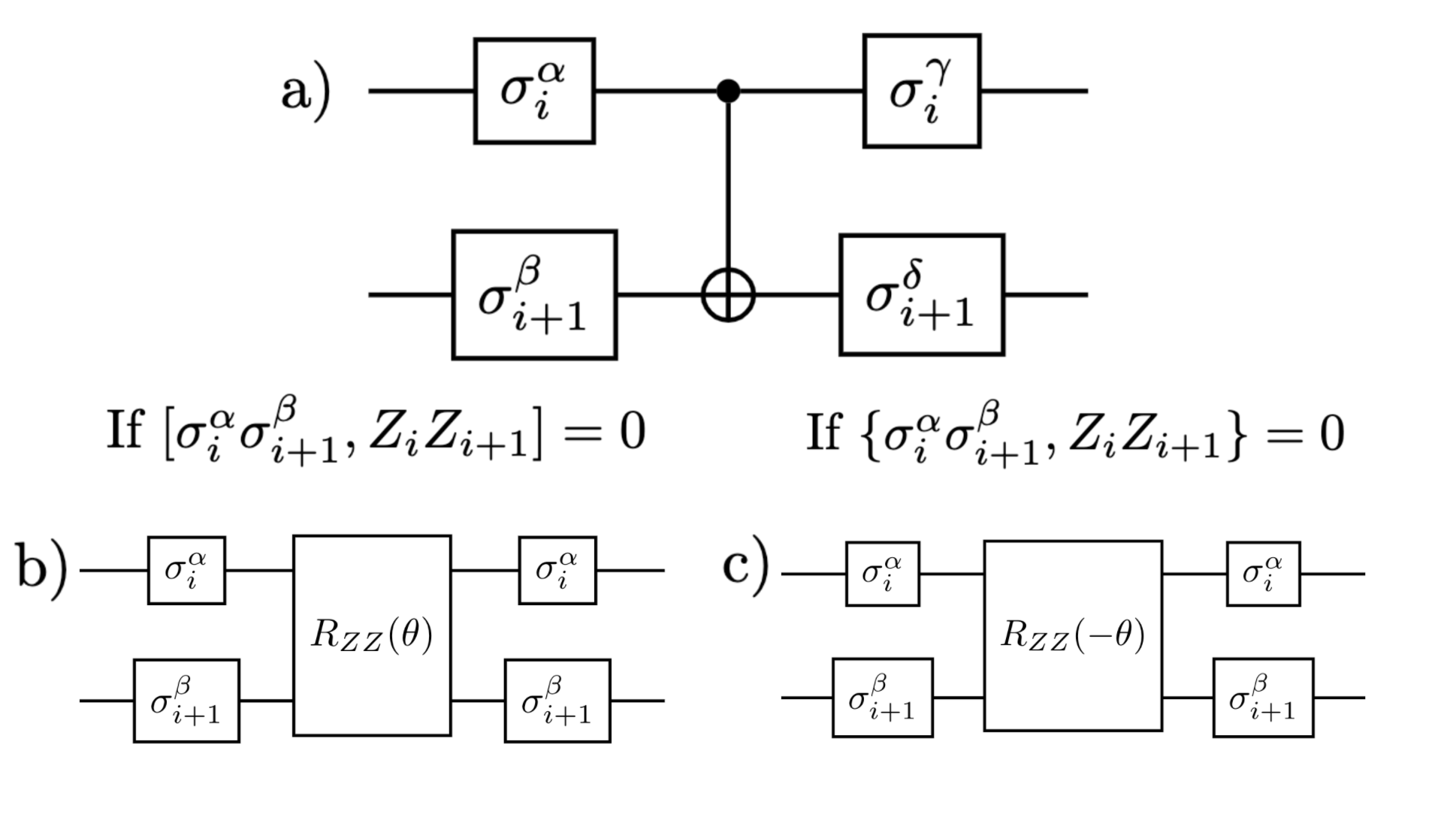}
\centering
\caption{
Implementation of Pauli twirling for (a) a CNOT gate and (b) the non-Clifford $R_{ZZ}(\theta)$ gate. In both cases, the Pauli gates $\sigma_c^{\alpha}$ and $\sigma_t^{\beta}$ applied to the control and target qubits before the gate being twirled are chosen randomly. Pauli gates applied after the two-qubit gate being twirled are chosen such that the logical action of the two-qubit gate is unaffected. For the CNOT gate, $\gamma$ and $\delta$ are chosen as $\gamma=\alpha+\beta(\beta-1)(\frac{7}{2}-\beta)(1-\frac{2}{3}\alpha)$ and $\delta=\beta+\alpha(\alpha-3)(\beta\text{mod}2-\frac{1}{2})$. For the $R_{ZZ}(\theta)$ gate, the same Pauli gates are applied before and after $R_{ZZ}(\theta)$, but the rotation angle $\theta\to-\theta$ if the randomly selected Pauli gates anticommute with $\sigma^z_c \sigma^z_t$.
}
\label{fig:PT}
\end{figure}

Moreover, we also apply the Pauli twirling technique which converts a two qubit gate $G$'s errors into a stochastic form characterized by the effective noise superoperator~\cite{Li17} 
\begin{align} 
\label{eq:GSuper}
\bar{\mathcal{N}}_{G}=F_{G}\left[\mathbbm1\right]+\sum_{\left(\alpha,\beta\right)\neq\left(0.0\right)}\epsilon_{\alpha,\beta}\left[\sigma_{c}^{\alpha}\sigma_{t}^{\beta}\right].
\end{align}
where $F_{G}$ is the fidelity, $\sigma_{c}^{\alpha}$ and $\sigma_{t}^{\beta}$ ($\alpha,\beta=0,1,2,3$ correspond to Pauli matrices $\mathbbm1,\sigma^{x},\sigma^{y},\sigma^{z}$, respectively) are Pauli operators acting on a control qubit $c$ and a target qubit $t$, and $\epsilon_{\alpha,\beta}$ are error probabilities. The quantity $[\sigma_{c}^{\alpha}\sigma_{t}^{\beta}]$ is a superoperator that acts on a quantum state with density matrix $\rho$ as $[\sigma_{c}^{\alpha}\sigma_{t}^{\beta}]\rho=\sigma_{c}^{\alpha}\sigma_{t}^{\beta}\rho\sigma_{c}^{\alpha}\sigma_{t}^{\beta}$. To convert the error into this stochastic form, one sandwiches the two-qubit gate $G$ with Pauli gates $\sigma_{c}^{\alpha}\sigma_{t}^{\beta}$ and $\sigma^{\gamma}_{c}\sigma_{t}^{\delta}$ with $\alpha,\beta=0,1,2,3$ and $\gamma,\delta$ chosen such that $\sigma_{c}^{\gamma}\sigma_{t}^{\delta}=G^\dagger\sigma_{c}^{\alpha}\sigma_{t}^{\beta}G$. This ensures that the sandwiched operator $\sigma_{c}^{\alpha}\sigma_{t}^{\beta} G \sigma_{c}^{\gamma}\sigma_{t}^{\delta} = G$. Note that choosing $\gamma$ and $\delta$ in this way is possible only if $G$ preserves the Pauli group, i.e., if $G$ is a Clifford gate. After sandwiching, one averages the true noise superoperator over the two-qubit Pauli group [i.e., over all 16 possible pairs $(\alpha,\beta)$] to obtain Eq.~\eqref{eq:GSuper}. In practice, it is sufficient to generate circuits with randomly chosen $(\alpha,\beta)$ for each two-qubit gate, compute the output of each circuit, and average the result over as many randomly generated circuits as possible.

In this paper, we use two kinds of two-qubit gates, namely CNOT and $R_{ZZ}(\theta)$. For the CNOT gate, we randomly select $\alpha,\beta=0,1,2,3$ and choose $(\gamma,\delta)$ so that the random Pauli gates do not affect the CNOT operation, as described in Fig.~\ref{fig:PT}(a). As discussed above, choosing $\gamma$ and $\delta$ in this way is only possible because CNOT is a Clifford gate. However, the $R_{ZZ}(\theta)$ gate is a non-Clifford gate for generic values of $\theta\neq0,\pi$. 
To solve this problem, we divide the set of Pauli index pairs $\{(\alpha,\beta)\}$ into two sets: $S_{C}$, containing index pairs corresponding to two-qubit Pauli strings that commute with $\sigma^3_c\sigma^3_t$, and $S_{A}$, containing index pairs corresponding two-qubit Pauli strings that anticommute with $\sigma^3_c\sigma^3_t$. If the randomly selected pair $(\alpha,\beta)\in S_{C}$, we replace $R_{ZZ}(\theta)$ by $\sigma^\alpha_{c}\sigma^\beta_{t}R_{ZZ}(\theta)\sigma^\alpha_{c}\sigma^\beta_{t}$. If the pair $(\alpha,\beta)\in S_A$, we replace $R_{ZZ}(\theta)$ by $\sigma^\alpha_{c}\sigma^\beta_{t}R_{ZZ}(-\theta)\sigma^\alpha_{c}\sigma^\beta_{t}$. The circuits resulting from this procedure are shown in Fig.~\ref{fig:PT}. Note that this modified Pauli twirling procedure assumes that $R_{ZZ}(\theta)$ and $R_{ZZ}(-\theta)$ have the same gate error channel, which we denote by the noise superoperator $\mathcal N_{R_{ZZ}}$.

To verify that the non-Clifford Pauli twirling procedure described above still results in a stochastic error channel for $R_{ZZ}(\theta)$, we assume that the action of the noisy $R_{ZZ}(\theta)$ gate can be expressed in superoperator form as $\mathcal{N}_{R_{ZZ}}\mathcal{U}_{R_{ZZ}}$. Here, $\mathcal{U}_{R_{ZZ}}$ is a superoperator that acts as $\mathcal{U}_{R_{ZZ}}\rho=R_{ZZ}(\theta)\rho R_{ZZ}(-\theta)$, and the noise superoperator $\mathcal{N}_{R_{ZZ}}$ is expressed in Kraus form as $\mathcal{N}_{R_{ZZ}}\rho=\sum_{h}E_{h}\rho E_{h}^{\dagger}$ with Kraus operators $E_h=\sum_{\alpha,\beta} a_{h; \alpha,\beta} \sigma_c^{\alpha} \sigma_t^{\beta}$ satisfying $ \sum_{h} E_h E_h^{\dagger}=\mathbbm1$. The action of the twirled noisy $R_{ZZ}(\theta)$ gate on a state $\rho$ can then be written as
\begin{widetext}
\begin{align} 
\frac{1}{16}
\left\{ 
\sum_{\left( \alpha,\beta\right) \in S_{C}}
[\sigma_{c}^{\alpha}\sigma_{t}^{\beta}]
\mathcal{N}_{R_{ZZ}}
\mathcal{U}_{R_{ZZ}}
[\sigma_{c}^{\alpha}\sigma_{t}^{\beta}]
\rho
+
\sum_{\left( \alpha,\beta\right) \in S_{A}}
[\sigma_{c}^{\alpha}\sigma_{t}^{\beta}]
\mathcal{N}_{R_{ZZ}}
\mathcal{U}_{R_{ZZ}}^{-1}
[\sigma_{c}^{\alpha}\sigma_{t}^{\beta}]
\rho
\right\}
\equiv
\mathcal{\bar{N}}_{R_{ZZ}}
\mathcal{U}_{R_{ZZ}}\rho .
\end{align}
and $\mathcal{\bar{N}}_{R_{ZZ}}$ is the effective noise of $R_{ZZ}(\theta)$ after twirling. Since two-qubit Pauli strings with $(\alpha,\beta)\in S_{C}$ commute with $R_{ZZ}(\theta)$, the first term can be written as 
\begin{align*} 
\sum_{\left( \alpha,\beta\right) \in S_{C}}
[\sigma_{c}^{\alpha}\sigma_{t}^{\beta}]
\mathcal{N}_{R_{ZZ}}
\mathcal{U}_{R_{ZZ}}
[\sigma_{c}^{\alpha}\sigma_{t}^{\beta}]
\rho
=
\sum_{\left( \alpha,\beta\right) \in S_{C}}[\sigma_{c}^{\alpha}\sigma_{t}^{\beta}]\mathcal{N}_{R_{ZZ}}[\sigma_{c}^{\alpha}\sigma_{t}^{\beta}]
\mathcal{U}_{R_{ZZ}}
\rho.
\end{align*}
Since two-qubit Pauli strings with $(\alpha,\beta)\in S_{A}$ anticommute with $R_{zz}(\theta)$, the second term becomes
\begin{align*} 
\sum_{\left( \alpha,\beta\right) \in S_{A}}
[\sigma_{c}^{\alpha}\sigma_{t}^{\beta}]
\mathcal{N}_{R_{ZZ}}
\mathcal{U}_{R_{ZZ}}^{-1}
[\sigma_{c}^{\alpha}\sigma_{t}^{\beta}]
\rho
=
\sum_{\left( \alpha,\beta\right) \in S_{A}}
[\sigma_{c}^{\alpha}\sigma_{t}^{\beta}]
\mathcal{N}_{R_{ZZ}}
[\sigma_{c}^{\alpha}\sigma_{t}^{\beta}]
\mathcal{U}_{R_{ZZ}}
\rho.
\end{align*}
\end{widetext}
Summing up two terms above, we find that the effective noise superoperator of $R_{zz}(\theta)$ is given by
\begin{align} 
\mathcal{\bar{N}}_{R_{ZZ}}
=
\frac{1}{16}
\sum_{\left( \alpha,\beta\right) }
[\sigma_{c}^{\alpha}\sigma_{t}^{\beta}]
\mathcal{N}_{R_{ZZ}}
[\sigma_{c}^{\alpha}\sigma_{t}^{\beta}].
\end{align}
Using $\sigma^{\alpha}\sigma^{\beta}\sigma^{\alpha}=[2\delta_{\alpha,\beta}-(2\delta_{\alpha,0}-1)(2\delta_{\beta,0}-1)]\sigma^{\beta}$, one can show that this effective noise superoperator takes the stochastic form~\eqref{eq:GSuper}. This modified Pauli twirling procedure can also be applied to other two-qubit unitary gates generated by Pauli strings. 

\subsection{Dynamical Decoupling}
In a quantum computer, physical two-qubit gates have different execution times. When we stack one- and two-qubit gates into a Trotter circuit, there are some idle qubits suffering from thermal relaxation and white noise dephasing. To reduce the decoherence, one can apply appropriate pulse sequences to stabilize the idle qubits during this waiting period. Here, we utilize the pulse sequence $\tau_{\rm iq}/4-X_{ \pi}-\tau_{\rm iq}/2-X_{- \pi}-\tau_{\rm iq}/4$ with $\pm \pi$ pulse $X_{\pm \pi}=R_{X}(\pm \pi)$ and delay time $\tau_{\rm iq}=\left(T_{\text{idle}}-2t_{x,\pi}\right)$. Here, $T_{\text{idle}}$ is the idle time of the qubit and $t_{x,\pi}$ is the duration of the $X_{\pm \pi}$ pulse~\cite{Lorenza98,Pokharel18,Jurcevic2021}. 

\section{Site-dependent local magnetization results}
\label{app:Qubits}

\begin{figure*}[h]
\includegraphics[width=2.0\columnwidth]{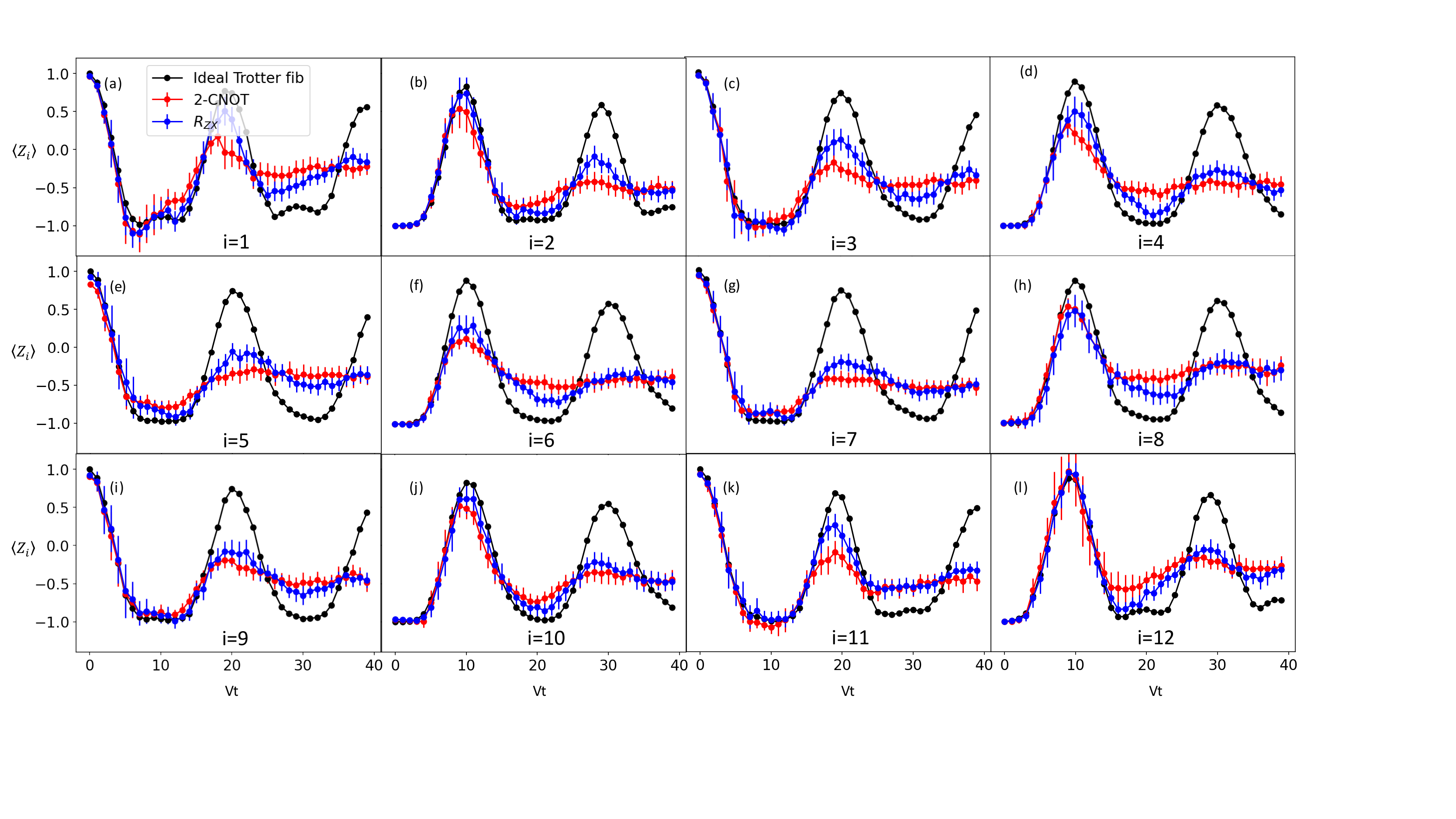}
\centering
\caption{Complete list of error-mitigated local magnetization results $\langle Z_i \rangle$ versus time $Vt$ for a 12-site chain measured on \texttt{ibmq\_guadalupe} using the scaled-$R_{ZX}$ and two-CNOT implementations (blue and red, respectively). The ideal Trotter simulation data (black) are also shown for reference.}
\label{fig:Z_I_12}
\end{figure*}

\begin{figure*}[h]
\includegraphics[width=2.0\columnwidth]{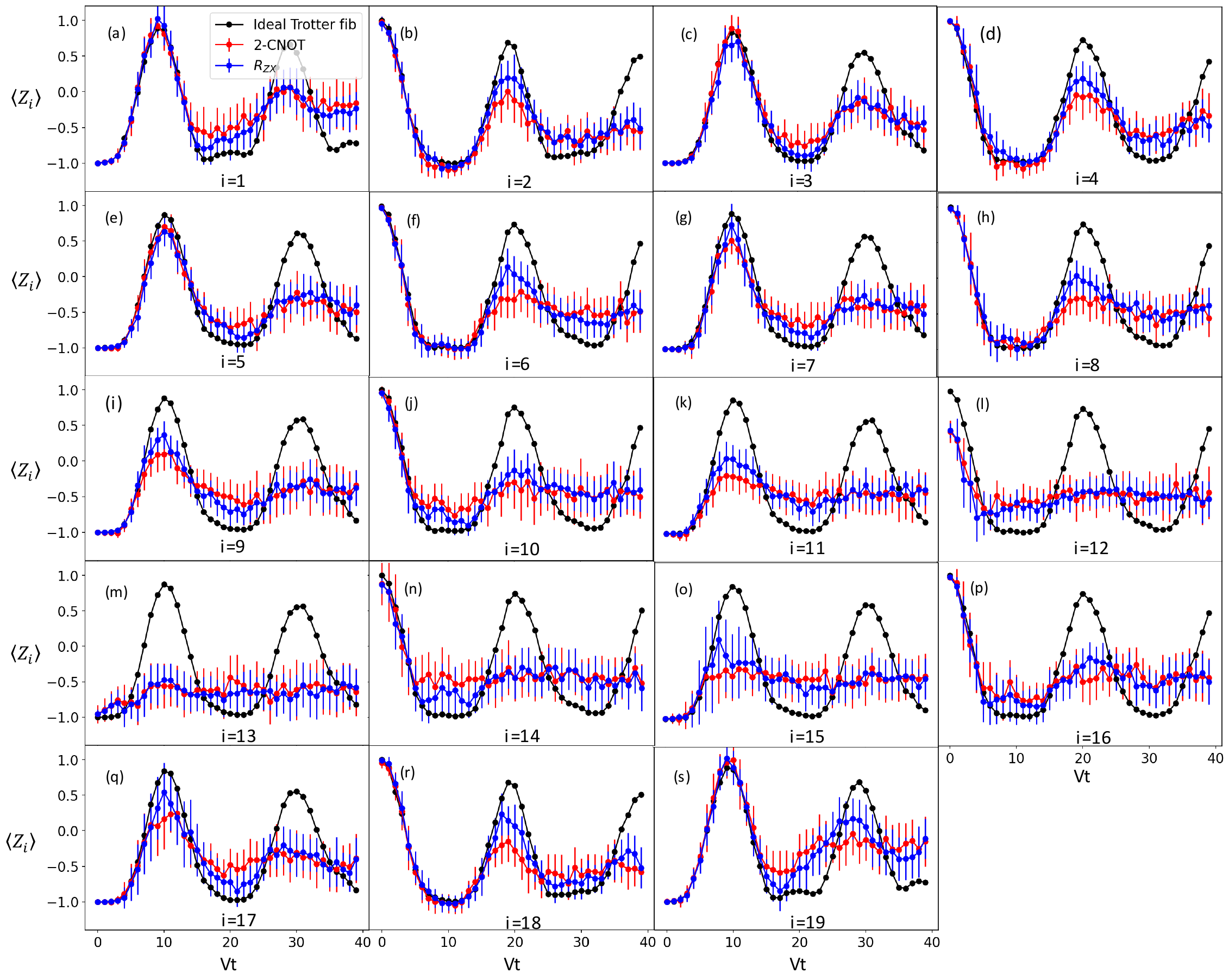}
\centering
\caption{Complete list of error-mitigated local magnetization results $\langle Z_i \rangle$ versus time $Vt$ for a 19-site chain measured on \texttt{ibmq\_toronto} using the scaled-$R_{ZX}$ and two-CNOT implementations (blue and red, respectively). The ideal Trotter simulation data (black) are also shown for reference.}
\label{fig:Z_I_19}
\end{figure*}

In this Appendix we show error-mitigated results for the site-wise local magnetization $\braket{Z_i (t)}$ as a function of scaled simulation time $Vt$ for a 12-site chain in Fig.~\ref{fig:Z_I_12} and a 19-site chain in Fig.~\ref{fig:Z_I_19}. The calculations are carried out on \texttt{ibmq\_guadalupe} and \texttt{ibm\_toronto}, respectively. We have use the full set of error mitigation techniques described in the main text and Appendix~\ref{app:Mitigation}.  Consistent with other observables discussed in the main text, the local magnetizations measured with the scaled-$R_{ZX}$ implementation on QPU are generally in better agreement with the ideal Trotter simulations than the results from the two-CNOT implementation. As an example, in Fig.~\ref{fig:Z_I_12}(e), the oscillatory behavior of $\braket{Z_5(t)}$ for the 12-site model in the second oscillation cycle is still visible with the scaled-$R_{ZX}$ implementation, but is completely washed out by noise with the two-CNOT implementation on \texttt{ibmq\_guadalupe}. The accuracy of local magnetization measurement also shows clear site-dependence, tied to the heterogeneity of qubit quality and native gate fidelity. For example, with the scaled-$R_{ZX}$ implementation in the 12-site model, an oscillation for two cycles can be clearly observed for $\braket{Z_2(t)}$, while $\braket{Z_6(t)}$ shows only a weaker first period of oscillation, as shown in Fig.~\ref{fig:Z_I_12}(b) and (f). The site-dependence of local magnetization accuracy becomes more evident for the 19-site model calculations on \texttt{ibmq\_toronto}. For instance, while the oscillation of $\braket{Z_{3}(t)}$ is still well reproduced over two cycles, $\braket{Z_{13}(t)}$ is almost entirely dominated by noise as shown in Fig.~\ref{fig:Z_I_19}(c) and (m).

\end{appendix}

\clearpage

\bibliography{Refs}

\end{document}